%% file: crrnc_new.tex
\documentclass{cimento}

\usepackage{graphicx,ctable}  

 \usepackage{lscape}
 
\title{Domenico Pacini,\\uncredited pioneer of the discovery of cosmic rays}
\author{A.~de Angelis\from{ins:ud} 
}
\instlist{\inst{ins:ud} Max-Planck-Institut f\"ur Physik
(Werner-Heisenberg-Institut),
F\"ohringer Ring 6,
M\"unchen, Germany, \thanks{On leave of absence from Universit\`a di Udine, Via delle Scienze 208, Udine, Italy.}; INFN and INAF Trieste, Italy; LIP/IST, Lisboa, Portugal
}

\PACSes{
\PACSit{96.50.S}{Cosmic rays}
\PACSit{96.50.sd}{Extensive air showers} 
\PACSit{92.60.Pw}{Atmospheric electricity}
\PACSit{01.65.+g}{History of Science}
}

\begin{document}

\maketitle

\begin{abstract}

During a series of experiments performed between 1907 and 1911, the Italian physicist Domenico Pacini (Marino 1878 -- Roma 1934), at that time researcher at the Central Bureau of Meteorology and Geodynamics in Roma, studied the origin of the radiation today called ``cosmic rays'', the nature of which was unknown at that time.  In his conclusive measurements in June 1911 at the Naval Academy in Livorno, and confirmed in Bracciano a couple of months later, Pacini,
proposing a novel experimental technique, observed the radiation strength to decrease when going from the surface to a few meters underwater (both in the sea and in the lake), thus demonstrating that such radiation could not come from the Earth. Pacini's conclusive experiment was performed, and the results published (in Italian), one year before the famous balloon experiment by Victor Hess, who found the ionization rate to increase with height.  While Hess is today celebrated as the discoverer of cosmic rays, Pacini's work was largely overlooked.  Hess was awarded the Nobel Prize in Physics in 1936, two years after the death of Pacini, who had become a full professor of Experimental Physics at the University of Bari and the Director of the local Institute of Physics. 
The discovery of cosmic rays - a milestone in science - involved several scientists in Europe and in the United States of America and took place during a period characterized by nationalism and a lack of communication.  Historical, political and personal facts, embedded in the pre- and post-World War I historical context, might have contributed to the substantial disappearance of Pacini from the history of science.  This article aims to give an unbiased historical account of the discovery of cosmic rays; in the centenary of Pacini's pioneering experiments, his work, which employed a technique that was complementary to, and independent from that of Hess, will be duly taken into consideration.  A translation into English of three fundamental early articles by Pacini is provided in the Appendix.

\end{abstract}

\tableofcontents

\section[A short history of the study of cosmic rays]{A short history of the study of cosmic rays\footnote{This Section is taken almost verbatim from \cite{noi2}, with the permission of the Authors.}}

Already in 1785 de Coulomb found \cite{Cou1785} that electroscopes can spontaneously discharge by the action of the air and not by defective insulation. After dedicated studies by Faraday around 1835 \cite{Far1835}, Crookes observed in 1879 \cite{Cro1879} that the speed of discharge of an electroscope decreased when pressure was reduced. The explanation of the phenomenon came in the beginning of the 20th century and paved the way to one of mankind's revolutionary scientific discoveries: cosmic rays. 
Several  reviews have been written on the history of the research on cosmic rays, see for example 
 \cite{montgo,janossy,leprince,rossi,hillas,wilson,wigandr,millikanr,comptonr,swann,xu,ginz,puppi,wolfe,casta}.

\subsection{The puzzle of atmospheric ionization}

Spontaneous radioactivity was discovered in 1896 by Becquerel \cite{Beq1896}. A few years later, Marie and Pierre Curie \cite{Cur1898} discovered that the elements Polonium and Radium suffered transmutations generating radioactivity: such transmutation processes were then called ``radioactive decays''. In the presence of a radioactive material, a charged electroscope promptly discharges. It was concluded that some elements were able to emit charged particles, that in turn caused the discharge of the electroscopes.  The discharge rate was then used to gauge the level of radioactivity. 

Following the discovery of radioactivity, a new era of research in discharge physics was then opened, this period being strongly influenced by the discoveries of the electron and of positive ions. During the first decade of the 20th century results on the study of ionization phenomena came from several researchers in Europe and in the United States. 

Around 1900, Wilson \cite{Wil1901} and Elster and Geitel \cite{ElG1900} improved the technique for a careful insulation of electroscopes in a closed vessel, thus improving the sensitivity of the electroscope itself. As a result, they could make quantitative measurements of the rate of spontaneous discharge. They concluded that such a discharge was due to ionizing agents coming from outside the vessel. The obvious questions concerned the nature of such radiation, and whether it was of terrestrial or extra-terrestrial origin. The simplest hypothesis was that its origin was related to radioactive materials, hence terrestrial origin was a commonplace assumption. An experimental proof, however, seemed hard to achieve. Wilson \cite{Wil1901} tentatively made the visionary suggestion that the origin of such ionization could be an extremely penetrating extra-terrestrial radiation. However, his investigations in tunnels with solid rock overhead showed no reduction in ionization \cite{Wil1901} and could therefore not support an extra-terrestrial origin. An extra-terrestrial origin, though now and then discussed (e.g. by Richardson \cite{Ric1906}), was dropped for the following years.

In 1903 Rutherford and Cook \cite{RuC1903} and also McLennan and Burton  \cite{MLB1903} showed that the ionization was significantly reduced when the closed vessel was surrounded by shields of metal kept free from radioactive impurity. This showed that part of the radiation came from outside. Later investigations showed that the ionization in a closed vessel was due to a ``penetrating radiation'' partly from the walls of the vessel and partly from outside. 

The situation in 1909 is well summarized by Kurz \cite{kurz}  and by Cline \cite{Cline}.  The spontaneous discharge was consistent with the hypothesis that even in insulated environments a background radiation did exist. In the 1909 review by Kurz three possible sources for the penetrating radiation are discussed: an extra-terrestrial radiation possibly from the Sun \cite{Ric1906}, radioactivity from the crust of the Earth, and radioactivity in the atmosphere. Kurz concludes from the ionization measurements in the lower part of the atmosphere that an extra-terrestrial radiation is unlikely.
It was generally assumed that large part of the radiation came from radioactive material in the crust. Calculations were made of how the radiation should decrease with height (see, {\em{e.g.,}} Eve \cite{eve}) and measurements were performed.

\begin{figure}
\begin{center}
\resizebox{0.5\columnwidth}{!}{\includegraphics{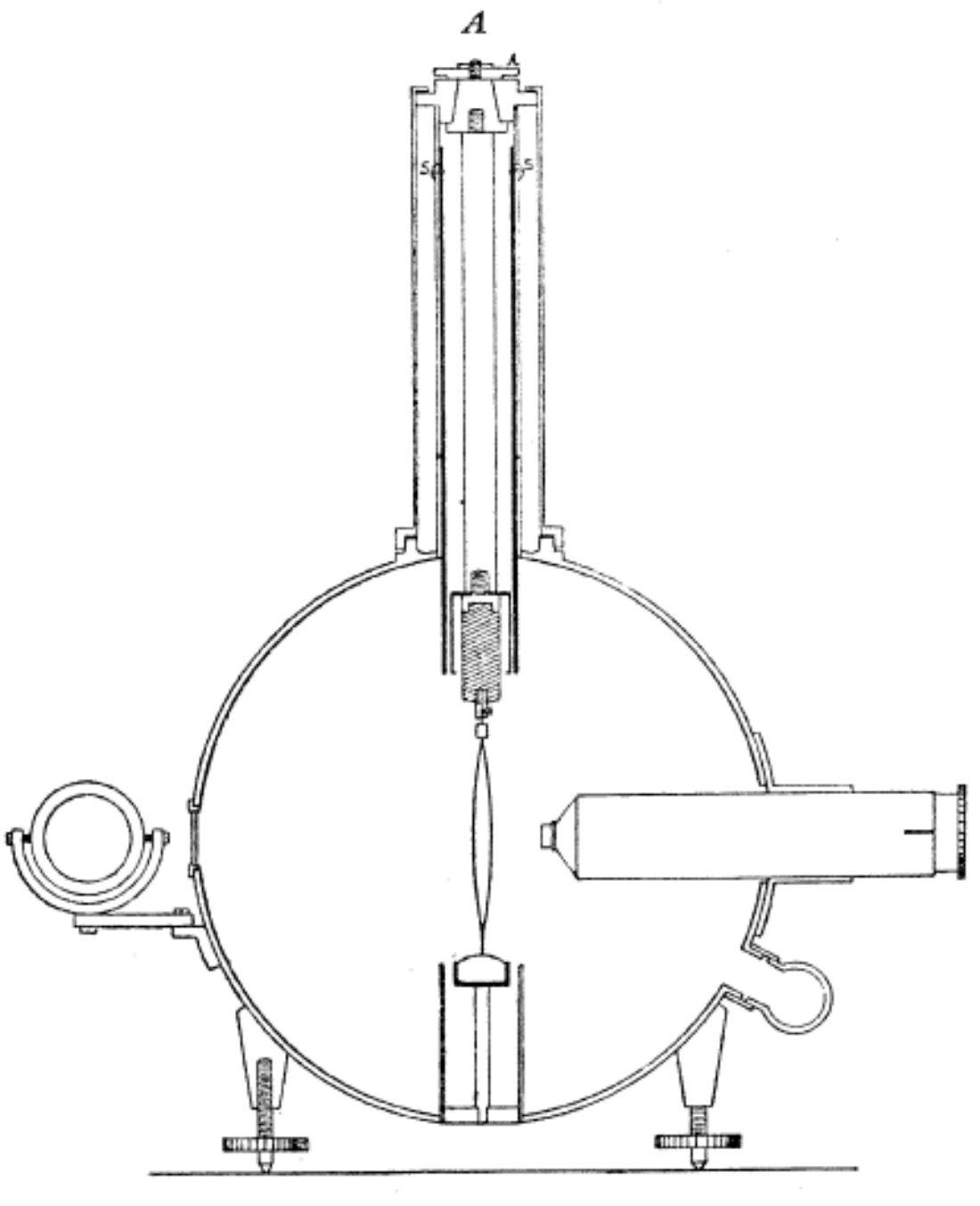} }
\end{center}
\caption{The Wulf electroscope. The 17 cm diameter cylinder with depth 13 cm was made
of Zinc. To the right is the microscope that measured the distance between the two silicon
glass wires illuminated using the mirror to the left. The air was kept dry using Sodium in the small container below the microscope. According to Wulf \cite{Wul1910}, with 1.6 ion pairs
per second produced, the tension was reduced by 1~V, the sensitivity of the instrument,
as measured by the decrease of the inter-wire distance.}
\label{fig:1}       
\end{figure}

Father Theodor Wulf, a German scientist and a Jesuit priest serving in the Netherlands and then in Roma ,  had the idea to check the variation of radioactivity with height to test its origin. In 1909 \cite{Wul1910} Wulf, using an improved  electroscope (Fig. \ref{fig:1}) in which the two leaves had been replaced by two metalized silicon glass wires, with a tension spring made also by glass in between, measured the rate of ionization at the top of the Eiffel Tower in Paris (300 m above ground). Supporting the hypothesis of the terrestrial origin of most of the radiation, he expected to find at the top  much  a smaller ionization than on the ground. The rate of ionization showed, however, too small a decrease to confirm the hypothesis. He concluded that, in comparison with the values on the ground, the intensity of radiation ``decreases at nearly 300 m [altitude] not even to half of its ground value''; while with the assumption that radiation emerges from the ground there would remain at the top of the tower ``just a few percent of the ground radiation'' \cite{Wul1910}. Wulf's observations were of great value, because he could take data at different hours of the day and for many days at the same place. For a long time, Wulf's data were considered as the most reliable source of information on the altitude effect in the penetrating radiation. However Wulf concluded that the most likely explanation of his puzzling result was still emission from the soil.

Other measurements with similar results were also made (Bergwitz \cite{Bergwitz}, McLennan and Macallum \cite{McLennan}, Gockel \cite{Goc1909}). The general interpretation of the outcome was that radioactivity was mostly coming from the Earth's crust.

\subsection{Pacini and the measurements of attenuation in water}

The conclusion that radioactivity was mostly coming from the Earth's crust was questioned by the Italian physicist Domenico Pacini, who compared the rate of ionization on mountains, over a lake, and over the sea \cite{Pac1909,Pac1910}; in 1911, he made important experiments by immersing an electroscope deep in the sea \cite{Pac1912}. 

\begin{figure}
\begin{center}
\resizebox{0.8\columnwidth}{!}{\includegraphics{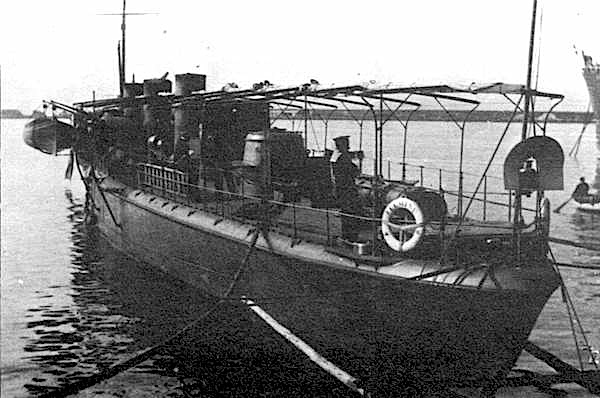} }
\end{center}
\caption{The cacciatorpediniere ``Fulmine'', used by Pacini for his measurements on the sea.}
\label{fig:fulmine}       
\end{figure}

Pacini, at that time an  assistant at Italy's Central Bureau of Meteorology and Geodynamics in Roma and later a professor in Bari, in a first period made several measurements to establish the variations of the electroscope's discharge rate as a function of the environment. First he placed the electroscope on the ground and on a sea a few kilometers off the coast. Pacini made his measurements over the sea in the Gulf of Genova, on an Italian Navy ship, the {\em cacciatorpediniere} (destroyer)  ``Fulmine" (Fig. \ref{fig:fulmine}) from the Accademia Navale di Livorno; the results were comparable. A summary of these results indicate, according to Pacini's conclusions, that ``in the hypothesis that the origin of penetrating radiations is in the soil, since one must admit that they are emitted at an almost constant rate (at least when the soil is not covered by remaining precipitations), it is not possible to explain the results obtained'' \cite{Pac1909}. Pacini's conclusion, confirmed by Gockel \cite{Goc1909}, was the first where it was established that the results of many experiments on radiation could not be explained by radioactivity in the Earth's crust.

Pacini continued the investigations of radiation and developed in 1911 a new experimental technique for the measurement of radioactivity a
few meters underwater\footnote{Mention of the Pacini experiments as the first example of ``under-surface'' measurements was made by C. Castagnoli \cite{casta}: ``[t]he depth was modest, but it established the beginning of a tradition of which we now represent the continuation.''};
 he found a significant decrease
by 20\% in the discharge rate when the electroscope was placed three meters underwater in the sea in front of the Naval Academy of Livorno (and later in the Lake of Bracciano), consistent with absorption by water of a radiation coming from outside.

Pacini reported those measurements, the ensuing results, and their interpretation, in a note titled {\em La radiazione penetrante alla superficie ed in seno alle acque (Penetrating radiation at the surface of and in water)} \cite{Pac1912}. 
 He concluded: ``[It] appears from the results of the work described in this Note that a sizable cause of ionization exists in the atmosphere, originating from penetrating radiation, independent of the direct action of radioactive substances in the soil." 

\subsection{Hess and the balloon measurements}
The need for balloon experiments \cite{libropd} became evident to clarify Wulf's observations on the effect of altitude (at that time and since 1885, balloon experiments were anyway widely used for studies of the atmospheric electricity). The first balloon flight with the purpose of studying the properties of penetrating radiation was arranged in Switzerland in December 1909 with a balloon called Gotthard from the Swiss aeroclub. Alfred Gockel, professor at the University of Fribourg, ascending up to 4500 m above sea level (a.s.l.) during three successive flights, found \cite{Goc1910,Goc1911} that the ionization did not decrease with height as expected on the hypothesis of a terrestrial origin. Gockel confirmed the conclusion of Pacini in \cite{Pac1909}, quoting him
correctly, and concluded ``that a non-negligible part of the penetrating radiation is independent of the direct action of the radioactive substances in the uppermost layers of the Earth'' \cite{Goc1911}. In a 1909 balloon ascent Bergwitz had found \cite{Bergwitz} that the ionization at 1300 m altitude had decreased to about 24\% of the value at ground, consistent with expectations if the radiation came from the EarthÕs surface. However, BergwitzÕ results were questioned because the electrometer was damaged during the flight (see, e.g., \cite{Goc1911}).

In spite of Pacini's conclusions, and of Wulf's and Gockel's puzzling results on the dependence of radioactivity on altitude, physicists were however reluctant to give up the hypothesis of a terrestrial origin. The situation was cleared up thanks to a long series of balloon flights by the Austrian physicist Victor Hess, who established the extra-terrestrial origin of at least part of the radiation causing the observed ionization. 


Hess, at that time working in Wien and in Graz, started his experiments by studying Wulf's results, and knowing the detailed predictions by Eve \cite{Eve1911} on the coefficients of absorption for radioactivity in the atmosphere. Eve wrote that, if one assumed a uniform distribution of RaC on the surface and in the uppermost layer of the Earth, ``an elevation to 100 m should reduce the [radiation] effect to 36 percent of the ground value''. Hess added: ``This is such a serious discrepancy [with Wulf's results] that its resolution appears to be of the highest importance for the radioactive theory of atmospheric electricity'' \cite{Hes1912}. Since in the interpretation of Wulf's and Gockel's results the absorption length of the radiation (at that time identified mostly with gamma radiation) in air entered crucially, Hess decided first to improve the experimental accuracy of the Eve's result by ``direct measurements of the absorption of gamma rays in air'' \cite{Hes1911}. He used  probes of about 1 g RaCl$_2$ at distances up to 90 m, and obtained an absorption coefficient consistent with Eve. Hence the contradiction of Wulf's results remained; Hess concluded that ``a clarification can only be expected from further measurements of the penetrating radiation in balloon ascents'' \cite{Hes1911}.

\begin{figure}
\begin{center}
\resizebox{0.6\columnwidth}{!}{ \includegraphics{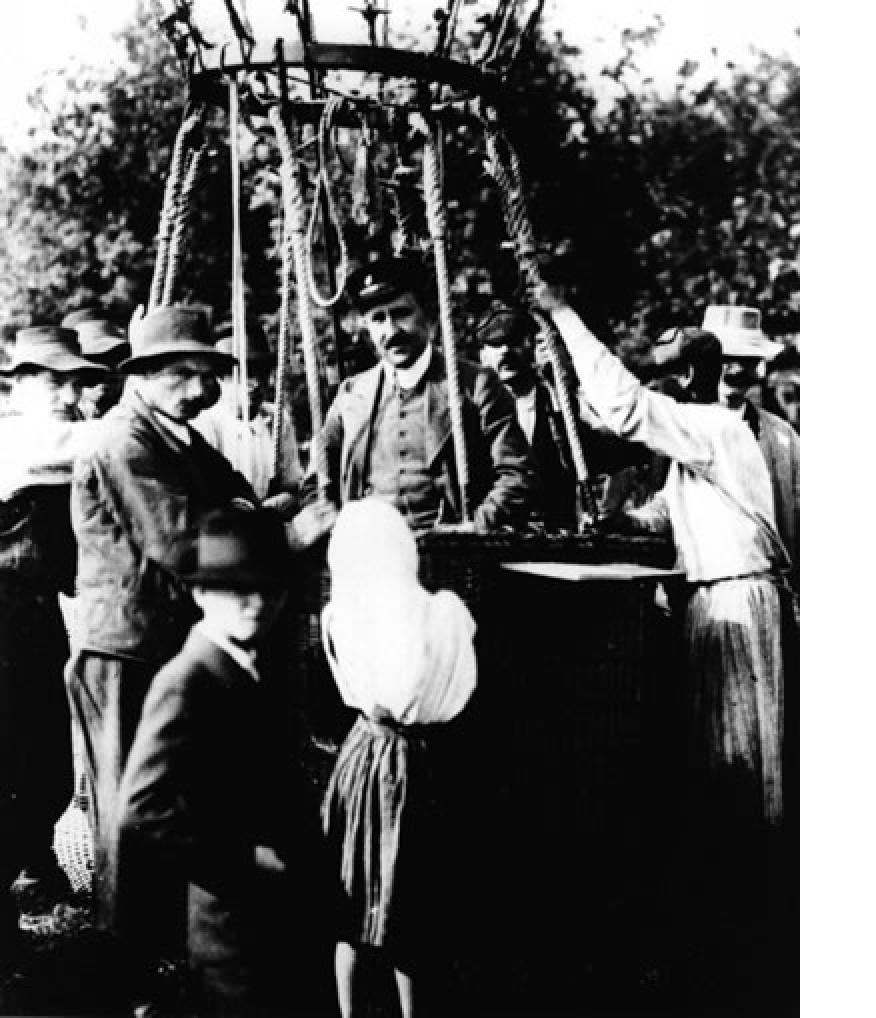} }
\end{center}
\caption{Historical photograph of Hess' balloon flight.}
\label{fig:balloon}       
\end{figure}

Hess continued his studies with balloon observations (Fig. \ref{fig:balloon}). The first ascension took place on August 28, 1911. ``[T]he balloon `Radetzky' of the Austrian aeroclub É with Oberleutnant S. Heller as pilot and me as sole passenger was lifted'' \cite{Hes1911}. The ascension lasted four hours and went up to a height of 1070 m above ground. A second ascension was done using another balloon (`Austria') during the night of 12 October 1911. During both balloon flights, the intensity of the penetrating radiation was measured to be constant with altitude within errors.

\begin{figure}
\begin{center}
\resizebox{\columnwidth}{!}{  \includegraphics{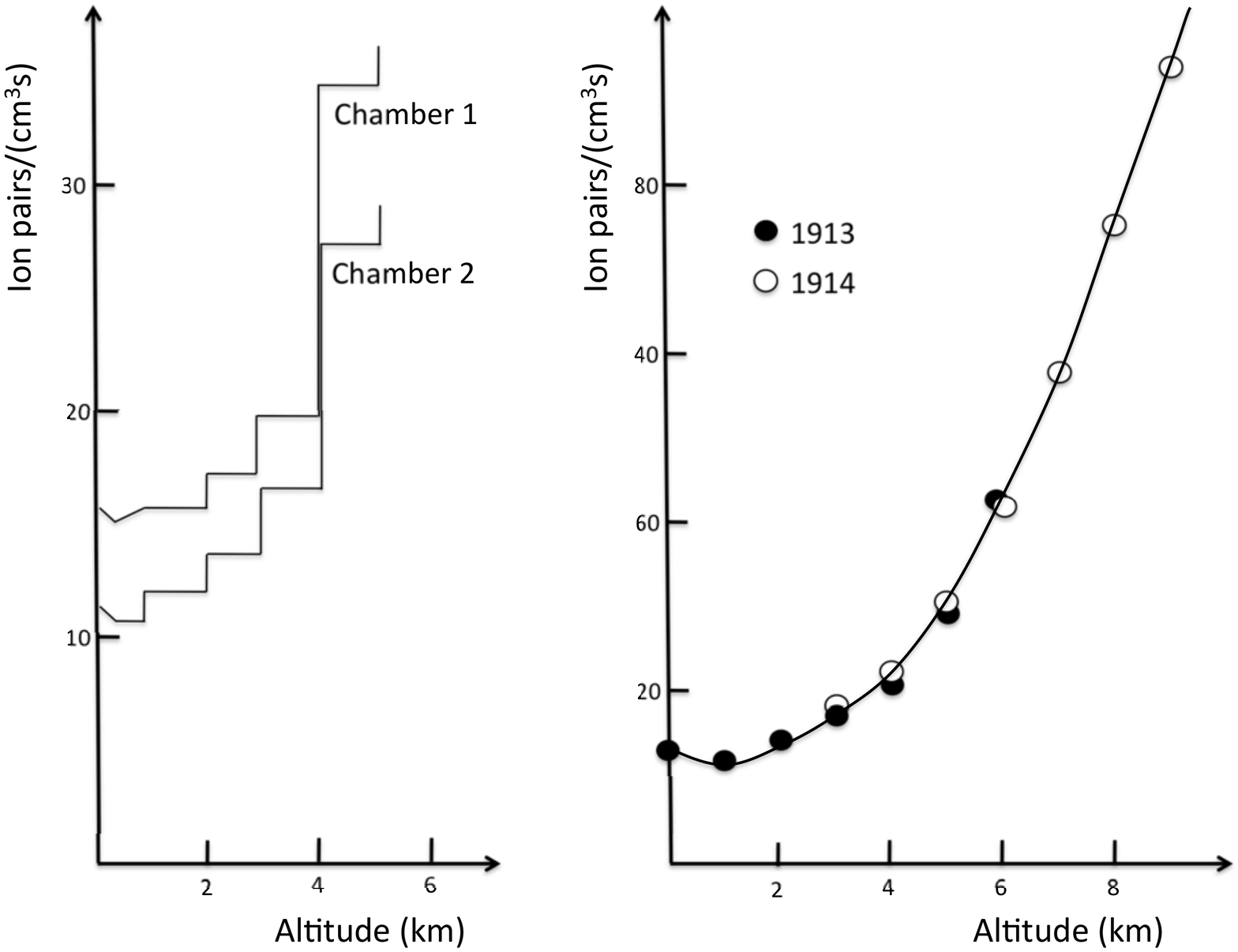} }
\end{center}
\caption{Variation of ionization with altitude. Left panel: Final ascent by Hess (1912), carrying two ion chambers (chamber 2 was shielded with thicker walls). Right panel: Ascents by Kolh\"orster (1913, 1914).}
\label{fig:increase}       
\end{figure}

From April 1912 to August 1912 Hess had the opportunity to fly seven times with three different instruments (enclosed in boxes with metal walls of
different thicknesses in order  to disentangle the effect of beta radiation). In the final flight, on August 7, 1912, he reached 5200 m. His results clearly showed that the ionization, after passing through a minimum, increased considerably with height (Fig. \ref{fig:increase}).  ``(i) Immediately above ground the total radiation decreases a little. (ii) At altitudes of 1000 to 2000 m there occurs again a noticeable growth of penetrating radiation. (iii) The increase reaches, at altitudes of 3000 to 4000 m, already 50\% of the total radiation observed on the ground. (iv) At 4000 to 5200 m the radiation is stronger [more than 100\%] than on the ground'' \cite{Hes1912}.

Hess concluded that the increase of the ionization with height must be due to a radiation coming from above, and he thought that this radiation was of extra-terrestrial origin. He also excluded the Sun as the direct source of this hypothetical penetrating radiation because of no day-night variation. Hess finally published a summary of his results in Physikalische Zeitschrift in 1913 \cite{Hes1913}, a paper which reached the wide public.

The results by Hess were later confirmed by Kolh\"orster \cite{Kol1914} in a number of flights up to 9200 m.  An increase of the ionization up to ten times that at sea level was found. The absorption coefficient of the radiation was estimated to be 10$^{-5}$ per cm of air at NTP. This value caused great surprise as it was eight times smaller than the absorption coefficient of air for gamma rays as known at the time.

\subsection{Developments after the World War I; discoveries in the US}

During the World War I (1914 - 1918) and the first years thereafter very few investigations of the penetrating radiation were performed.  Kolh\"orster continued his investigations using newly constructed apparatuses and made measurements at mountain altitudes with results published in 1923 in agreement with earlier balloon flights. There were, however, also negative attitudes in Europe against an extra-terrestrial radiation. Hoffmann, using newly developed electrometers, concluded \cite{hoff} that the cause of the ionization was radioactive elements in the atmosphere. Similar conclusions were reached by Behounek \cite{beh}.

After the war, the focus of the research moved to the US. Millikan and Bowen \cite{Mil1923} developed a low mass (about 200 g) electrometer and ion chamber for unmanned balloon flights using data transmission technology developed during World War I. In balloon flights to 15,000 m in Texas they were surprised to find a radiation intensity not more than one-fourth the intensity reported by Hess and Kolh\"orster. They attributed this difference to a turnover in the intensity at higher altitude, being unaware that a geomagnetic effect existed between the measurement in Europe and Texas. Thus, Millikan believed that there was no extraterrestrial radiation. As reported to the American Physical Society in 1925 Millikan's statement was ``The whole of the penetrating radiation is of local origin''.

In 1926, however, Millikan and Cameron \cite{Cam1926} carried out absorption measurements of the radiation at various depths in lakes at high altitudes. Based upon the absorption coefficients and altitude dependence of the radiation, they concluded that the radiation was high energy gamma rays and that ``these rays shoot through space equally in all directions'' calling them ``cosmic rays''\footnote{Hess used the name H\"ohenstrahlung (radiation from top) after his 1912 flights. Several other names were used  to indicate the extraterrestrial radiation before ``cosmic rays'' was generally accepted: Ultrastrahlung, kosmische Strahlung, Ultra-X-Strahlung.}. Pacini's work was not quoted.

Millikan was handling with energy and skill the communication with the media, and in the US the discovery of cosmic rays became, according to the public opinion, a success of American science. Millikan argued that the radiations are ``generated by nuclear changes having energy values not far from [those that they recorded] in nebulous matter in space.'' Millikan then proclaimed that this cosmic radiation was the ``birth cries of atoms'' in our galaxy. His lectures drew considerable attention from, among others, Eddington and Jeans, who struggled unsuccessfully to describe processes that could account for Millikan's claim.

\subsection{Properties of the radiation}
It was generally believed that the cosmic radiation was gamma radiation because of its penetrating power (one should remember that the penetrating power of relativistic charged particles was not known at the time). Millikan had put forward the hypothesis that such gamma rays were produced when protons and electrons form helium nuclei in the interstellar space. 

A key experiment, which would decide on the nature of cosmic rays (and in particular if they were charged or neutral), was the measurement of the dependence of cosmic ray intensity on geomagnetic latitude. Important measurements were made in 1927 and 1928 by Clay \cite{clay} who, during two voyages between Java and Genova, found that the ionization increased with latitude. No such variation was expected if the radiation was a gamma radiation, but Clay did not draw a firm conclusion as to the nature of the cosmic radiation. Clay's work was disputed by Millikan.

With the introduction of the Geiger-M\"uller counter in 1928, a new era began and soon confirmation came that the cosmic radiation is indeed corpuscular. Kolh\"orster introduced the coincidence technique. Bothe and Kolh\"orster \cite{BK} concluded that the cosmic radiation is mainly or fully corpuscular, but still Millikan did not accept this view. 

In 1932 Compton carried out a world-wide survey to settle the dispute.  He then reported \cite{Com1933} that there was a latitude effect, that cosmic rays were charged particles and that Millikan was wrong.  Millikan attacked strongly Compton, but after repeating his experiment in 1933 he admitted that there was a latitude effect and that the cosmic rays must be (mostly) charged particles. In 1933, three independent experiments by Alvarez \& Compton
\cite{aecew}, Johnson \cite{jew}, Rossi \cite{rew} discovered that
more cosmic rays were coming from West than from East close to the
Equator: this is due to the interaction with the magnetic field of the
Earth, and it demonstrates that cosmic rays are mostly positively
charged particles. However, it would take until 1941 before it was established in an experiment by Schein \cite{sche} that cosmic rays were mostly protons.

Much attention was in the middle of the 1920's given to the question of a variation with time of the radiation. Many investigators had found such variations, but towards 1930 the general opinion was that no such variation existed. The work of Forbush led to the proof that the observed intensity of cosmic rays within the atmosphere of Earth varied with time \cite{Forbush1938}. 

\subsection{Cosmic rays and the beginning of particle physics}
%
%

Thanks to the development of cosmic ray physics, scientists knew then that astrophysical sources were providing very-high energy bullets entering the atmosphere. It was then obvious to investigate the nature of such bullets, and to use them as probes to investigate matter in detail, along the lines of the experiment made by Rutherford in 1900. Particle physics, the science of the fundamental constituents of matter, started with cosmic rays. 
Many fundamental discoveries were made. 

The first was the discovery of the positron, the first identified antiparticle. While watching the tracks of cosmic rays passing through his cloud chamber, Anderson in 1933 discovered antimatter in the form of the anti-electron, later called the positron \cite{positron}. 

Neddermeyer and Anderson discovered in 1937 in cosmic rays the elementary subatomic particle called the muon \cite{muon}. The positron and the muon were the first of series of subatomic particles discovered using cosmic rays. Later came the discovery of the first meson, the charged pion by Powell, Lattes and Occhialini in 1947 \cite{pion}. Also the discovery of strangeness \cite{kaon} was made thanks to cosmic rays. Particle physicists used cosmic rays as the main tool for their research, and the pioneering results of that field are due to cosmic rays.  This includes the beginning of the discovery of new particles
(the so-called ``particle zoo'') in the tracks 
of atmospheric emulsions and chambers, and was the motivation of unifying and simplifying ideas like the theory of quarks. A historical account is 
provided, e.g., in \cite{birth}.

World War II interrupted the study of high-energy physics through cosmic rays but provided new resources, both technical and political, for the study of elementary particles. Technical resources included advances in microwave electronics and the design of human-made particle accelerators, which allowed physicists to produce high energy particles in a controlled environment. By about 1950 elementary particle physics would have been dominated by accelerator physics, at least till the beginning of the nineties, when the exploration possible with the energies one can produce on Earth started showing signs of saturation.

\subsection{Cosmic-ray physics today}
Cosmic rays are today central to the new field of astroparticle physics, an interdisciplinary field between astrophysics, cosmology and particle physics. Astroparticle physics has grown considerably the last 20 years and many large projects are being performed, searching, {\em{e.g.,}} for the dark matter of the universe. Satellite projects like the Fermi gamma-ray space telescope \cite{fermi} and the Pamela magnetic spectrometer \cite{pamela} have given new cutting-edge results. The planned AMS-02 \cite{ams} mission will further extend this type of research.

The study of the showers induced in the atmosphere by the very highest energy cosmic rays, 70 years after the discovery of air showers by Rossi and Auger\footnote{In 1934 Bruno Rossi reported an observation of near-simultaneous discharges of two Geiger counters widely separated in a horizontal plane during a test of equipment. In his report on the experiment, Rossi \cite{rossi0} wrote ``...it seems that once in a while the recording equipment is struck by very extensive showers of particles, which causes coincidences between the counters, even placed at large distances from one another.'' In 1937 Pierre Auger, unaware of Rossi's earlier report, detected the same phenomenon and investigated it in some detail. He concluded \cite{augso} that extensive particle showers are generated by high-energy primary cosmic-ray particles that interact with air nuclei high in the atmosphere, initiating a cascade of secondary interactions that ultimately yield a shower of electrons, photons, and muons that reach ground level. Explanation of the mechanism of formation of those Extensive Air Showers (EAS) was formulated by G.T. Zatsepin after observations at the Pamir mountain \cite{zat}.
A general theory of nuclear cascade processes was then published in \cite{rose}.}, is providing with the large
 ground-based Pierre Auger Observatory \cite{auger}  fundamental knowledge of the  high-end spectrum of cosmic rays, in particular giving a marginal indication that the direction of extremely high energy cosmic rays (above a few joule per particle) might be correlated to the nuclei of galaxies outside the Milky Way \cite{augsc}. The ground-based very-high energy gamma Cherenkov telescope arrays H.E.S.S. \cite{HESS}, MAGIC \cite{MAGIC} and VERITAS \cite{VERITAS} are mapping the cosmic emitters of gamma rays (and thus, indirectly, of cosmic rays) in the TeV region, providing evidence that  supernova remnants in the galaxy are emitters of cosmic rays up to a few hundreds of TeV \cite{zwi}.

In view of the fundamental limitations of earthbound high-energy physics experiments, cosmic rays and cosmological sources again move into the focus of very-high-energy particle and gravitational physics. 


\section{An in-depth look to the contribution of Domenico Pacini}


Around 1910, the Austrian Victor Hess and  the Italian Domenico 
Pacini independently carried out two different, ingenious, and 
complementary research lines that would eventually clarify the origin of the yet 
mysterious ionizing background radiation.

Pacini made several measurements to establish the variations of an electroscope's 
discharge rate as a function of the environment: he placed the instrument on the 
ground, on the sea a few kilometers off the coast, and a few meters underwater. 
After measuring a small decrease of the radioactivity  on the sea surface with respect to the radioactivity on the ground, in June 1911
 he measured a significant decrease in the discharge rate when the electroscope was placed 
underwater. He could thus firmly conclude that a relevant part of the radioactivity was not coming from the Earth's crust \cite{Pac1912}.

Hess used a different approach, and he measured the variation of the discharge rate of the
electroscope at different heights; for this purpose, he made observations on balloons. The first ascensions took place between August and October 1911, and the results were not significant. 
From April 1912 to August 1912 Hess had the opportunity to fly seven times with three instruments (two with thick walls and one with thin walls, to disentangle the effect of beta radiation). In the final flight, on August 7, 1912, he reached 5200 m. His results clearly showed that the ionization, after passing through a minimum, increased considerably with height (Fig. \ref{fig:increase}).    This was a demonstration \cite{Hes1912} that a large part of the radiation was of extraterrestrial origin (what is called today ``cosmic rays'').

Who was Domenico Pacini \cite{Riz1934,rob,stra1,stra2} and, while Victor Hess is honored as the discoverer 
of cosmic rays, why did Pacini's  earlier discovery go unnoticed and 
was soon forgotten (notably in Italy)? Personal stories and historical events contributed 
to this outcome.

\subsection{Pacini: a short biography}

\begin{figure}
\begin{center}
\resizebox{0.5\columnwidth}{!}{\includegraphics{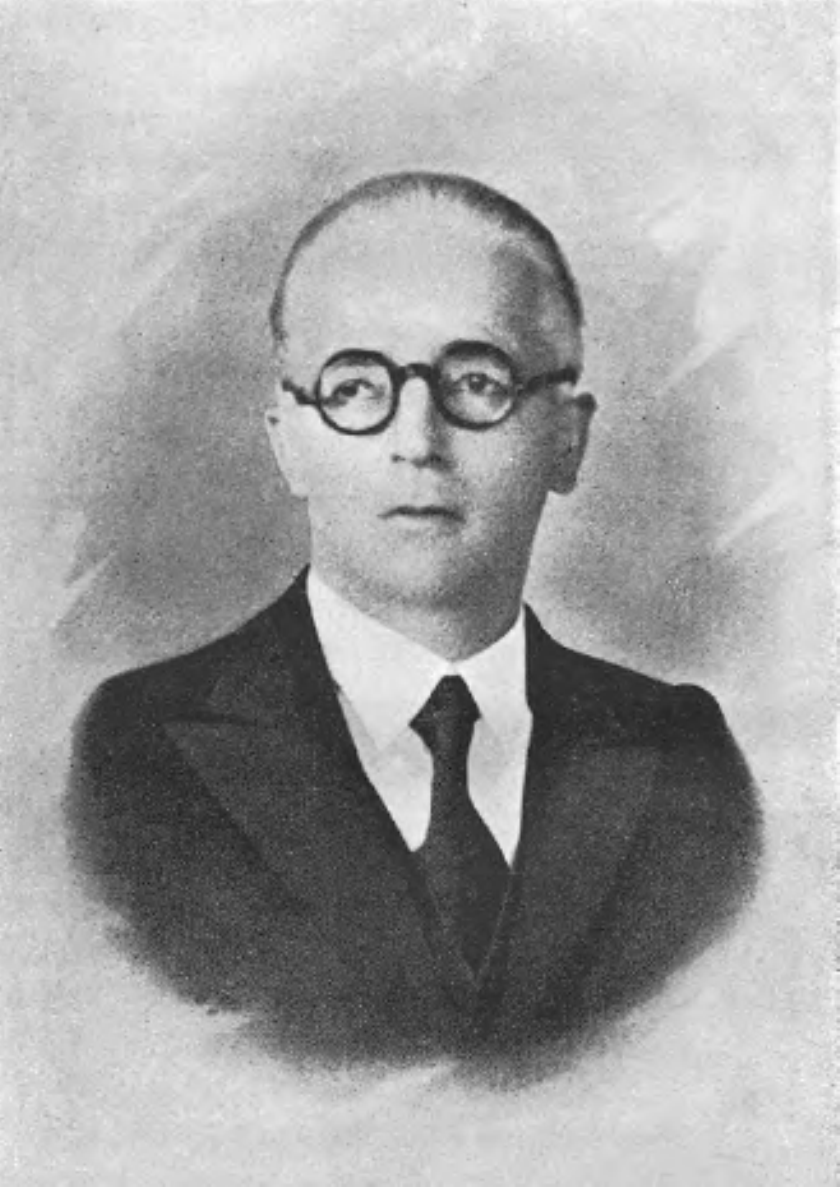} }
\end{center}
\caption{Domenico Pacini.}
\label{fig:2}       
\end{figure}

Domenico Pacini (Fig. \ref{fig:2}) was born on February 20, 1878, in Marino, near Roma, the son of Filippo, 30, {\em segretario comunale} (principal clerk of the city), who had studied  philosophy at the Pontifical University in Roma, and of Giovanna Mecheri, 21, housewife. In 1885 he moved with his family to Forme, a fraction of Massa d'Albe, near l'Aquila.

After technical studies at the high school ``Leonardo da Vinci'' in Roma, Domenico Pacini graduated in 
Physics in 1902 at the Faculty of Sciences of the Roma University. 

At the University of Roma, for the 
next three years, he worked as a junior assistant of Professor Pietro Blaserna\footnote{Pietro Blaserna (Fiumicello di Aquileia, 1836 - Roma, 1918) was a Professor of Experimental Physics at the University of Palermo, and, since 1879, President of the Italian Bureau of Meteorology and Geodynamics. Named senator in 1890, since 1906 he was vice-president of the Senate. Blaserna worked on electromagnetic induction, on the measurement of refractive indices, on the kinetics of gases, and on acoustic. He was among the founders of the Societ\`a Italiana di Fisica, and its first president in 1897 \cite{wiki,sif}.} while also 
studying electric conductivity in gaseous media under the supervision of Professor Alfonso 
Sella\footnote{Alfonso Sella (Biella, 1865 - Roma, 1907)
was the son of Quintino Sella, mineralogist and Prime minister of the Italian Government.
Professor at the University of Roma, he worked on radioactivity.}. 
In 1904 Pacini set out to study the infamous N-rays: he performed an 
experiment, the (null) results of which were communicated in a letter to Nature~\cite{nrays}
as ``careful experiments made ... with the object of observing the
effects of N-rays described by M. Blondlot and other investigators.''
Though ``observations were made under very
favourable conditions,'' he ``was unable to detect any increase
of luminosity of a phosphorescent screen caused by unknown
rays from strained or tempered steel, an Auer lamp, a
Nernst lamp, sound vibrations, or a magnetic field, though
various French observers have affirmed that in each of these
cases N-rays are emitted which produce an effect upon the
screen''. 

In August 1905, Pacini obtained a permanent position as an
assistant at Italy's {\em Regio Ufficio Centrale di Meteorologia e Geodinamica} (Central Bureau of Meteorology and Geodynamics), directed at that time
by Professor Luigi Palazzo,\footnote{Luigi Palazzo (Torino, 1861 - Firenze, 1933) was a geophyscist and an academician; he directed the Central Bureau of Meteorology and Geodynamics from 1900 to 1931~\cite{treccani}.}
working in the group  in charge of studying thunderstorms and electric phenomena 
in the atmosphere.  Many of the activities of the Central Bureau were requiring travels into different parts of Italy: Pacini spent long periods, in particular, in Castelfranco Veneto, between Padova and Treviso, and in the meteorological observatory of Sestola, near Modena, at an altitude of 1090 m above sea level.

In July 1913 Pacini obtained, on the proposal by a board chaired by Professor Vito
Volterra\footnote{Vito Volterra (Ancona, 1860 - Roma, 1940) was a mathematician and physicist, known for his contributions to mathematical biology and integral equations. He became professor of mathematical physics in Pisa in 1883 and then, in 1900, in the University of Roma. Named  as a senator in 1905, in 1922 he joined the opposition to the Fascist regime of Mussolini and in 1931 he was one of the few professors who refused to take a mandatory oath of loyalty. As a result he was compelled to resign his university post and his membership of scientific academies, and, during the following years, he lived largely abroad, returning to Roma just before his death. One of the founders of the {\em Societ\`a Italiana di Fisica,} he has been President of the
Societa\`a itself from 1909 to 1919, and President of the {\em Consiglio Nazionale delle Ricerche,} the Italian National Research Council.}, the {\em libera docenza} (habilitation to teach) in Experimental Physics. He was charged of the course of Mathematics for students of Chemistry and of Natural Sciences at the University of Roma from 1915 to 1919, and then of Earth's Physics from 1924 to 1925.

In 1926 Pacini participated to a selection for a full professorship in Experimental Physics at the University of Bari, and he was ranked third. 
In 1927 he was promoted to Principal Geophysicist at the Central Bureau of Meteorology and Geodynamics; however, since the two best ranked candidates, Polvani\footnote{Giovanni Polvani (Spoleto, 1892 - Milano, 1970) was an experimental physicist, President of the {\em Societ\`a Italiana di Fisica} and of the {\em Consiglio Nazionale delle Ricerche.}} and Rita Brunetti, had been nominated Professors in Pisa and in Ferrara respectively, Pacini was appointed full Professor at the University of Bari in 1928, and he resigned from the {\em Regio Ufficio Centrale di Meteorologia e Geodinamica}. In Bari he was incharged of setting up the studies of  Physics within the Faculty of Medicine, and of reorganizing the Physics Department, that he directed. He lectured on General Physics for Life Sciences, and on Mineralogy. While in Bari, 
his research interests mainly focused on the diffusion processes of light in the atmosphere. 
In 1932 Guglielmo Marconi selected him  in the National Committee for Astronomy, Applied Mathematics and Physics of the ``Consiglio Nazionale delle Ricerche''.

Domenico Pacini died of pneumonia in Roma on May 23, 1934,  one year after the death of his mother, and shortly after marrying with Mrs. Pierina Rangoni from Bologna, who donated all his manuscripts and books to the University of Bari. 

The City of Marino and the University of Bari independently organized solemn celebrations for Pacini, which were held in in July; Professor Giovanni Battista Rizzo from Napoli was the official biographer \cite{Riz1934} for the Societ\`a Italiana di Fisica.

Upon his death Pacini was buried in Roma's ``Cimitero del Verano''. In 1988, however, his remains were moved to the Cemetery of Forme,   where his family had established and where he had worked as a young researcher.


\subsection{The contribution of Pacini to research}

 The contribution of Domenico Pacini to physics before 1913, according to Pacini himself \cite{pacld}, was developed along three lines of research:
\begin{enumerate}
\item Solar radiation, terrestrial magnetism, and meteorology \cite{PacMet0,PacMet1,PacMet2,PacMet3} . The work by Pacini in this field appears to
be, according to the judgement of the board evaluating him for the habilitation to teach, mostly descriptive.
\item The electrical conductivity through gases  \cite{PacElectro1,PacElectro2,PacElectro3,PacElectro4,PacElectro5}. Related to this subject, one has to note that the conductivity
of air, as measured by the rate of discharge of an
electroscope, was studied as a part of atmospheric physics
(meteorology) at the beginning of the XX century \cite{xu}. This line of research allowed Pacini to improve his know-how on the instrumentation needed to measure low levels of ionization; this will be important for his main subject of research (item 3).
\item Radioactivity and atmospherical electricity. This is the main subject of Pacini's research, and it will be discussed in larger detail in the next Subsection.
\end{enumerate}
Besides the three topics listed above, after 1913 Pacini worked to the study of the emission of electric charge from radioactive salts \cite{sali}, to the spectroscopy of the light from the sky, in particular on the relation between the color of the sky and Avogadro's constant \cite{PaciniSpectro,PaciniSpectro1},
to the formation of condensation nuclei \cite{PaciniNuclei1,PaciniNuclei2,PaciniNuclei3}.

Related to the study of the color of the sky \cite{PaciniSpectro,PaciniSpectro1},
Pacini  made measurements in Sestola, on a hill 1090 m heigh, about the spectral composition of
the light scattered from the sky. He measured the ratio between the
intensity of the light scattered at 90$^\circ$ and the intensity of the light
emitted from the Sun for a number of wavelengths. In this way he determined  the
 product of the cross section times
the density of the molecules, and gave a rough
estimate of Avogadro's constant. Since part of the light of the sky is
due to the albedo from the Earth (the fraction of solar energy
reflected from the ground back into the atmosphere),  and
must be subtracted, Pacini was lead to measure the  albedo
itself in order to improve the quality of his results; these last
measurements were performed in Sestola and in Ciampino between 1927
and 1928 \cite{albedo}.



Related to the formation of condensation nuclei, Pacini was interested in the study of the condensation of steam over small impurities in the air, neutral or ionized. Pacini concluded  that this phenomenon happens when steam is over-saturated, at a level which becomes higher as condensation nuclei become smaller. Pacini proposed a technique to measure the density of particles of dust  from the rate of condensation.

Pacini wrote three review articles, two related to atmospherical electricity \cite{Pacr1910,Pacr1918}, and one on the phenomena of the high atmosphere \cite{Pacr1929}.

\subsection{The way that led Pacini to the hypothesis of extraterrestrial radiation}

Certainly the main contribution by Pacini to physics is related to the study of atmospherical ionization.

\begin{figure}
\begin{center}
\includegraphics[width=0.7\textwidth]{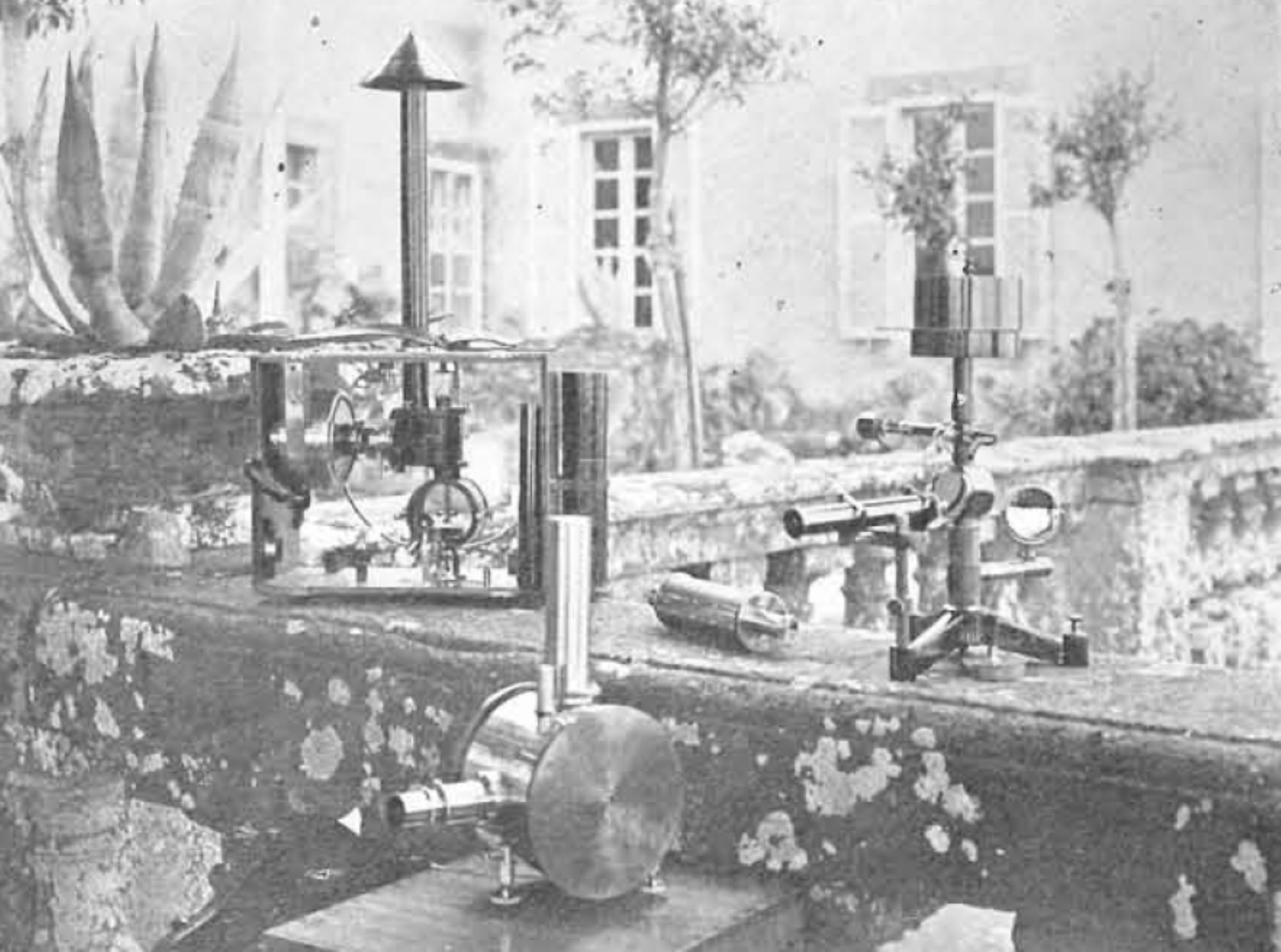}
\end{center}
\caption{Three different kinds of electroscopes actually used by Pacini for his measurements. In foreground, the Ebert's electroscope.}
\label{fig:elettroscopi}
\end{figure}

The long way that led Pacini to the hypothesis of cosmic rays (or, to be more precise, of radiation not coming from the Earth's crust)  started from his studies 
on electric conductivity in gaseous media that he performed at the University of Roma
during the early years of the XX century. While working at the Central Bureau of 
Meteorology, he became interested in the problem of the ionization of air, and technically acquainted with several instruments for the measurement of ionization (Fig. \ref{fig:elettroscopi}). Ionization can come from UV-rays, from wind, and from mechanical effects; however the main cause is due to radioactivity - although the final level of ionization is influenced by many effects, for example by humidity. Before studying in a proper way ionization, one needs thus a careful training to learn how to control systematic effects.

Pacini started systematic studies of ionization in 1905, and he demonstrated that the effect of ionization by X-rays is higher in the presence of some aerosols, like alcohol; he explained this fact as a more difficult recombination induced by the vapors \cite{PacElectro4}.

\begin{figure}
\begin{center}
\resizebox{0.75\columnwidth}{!}{ \includegraphics{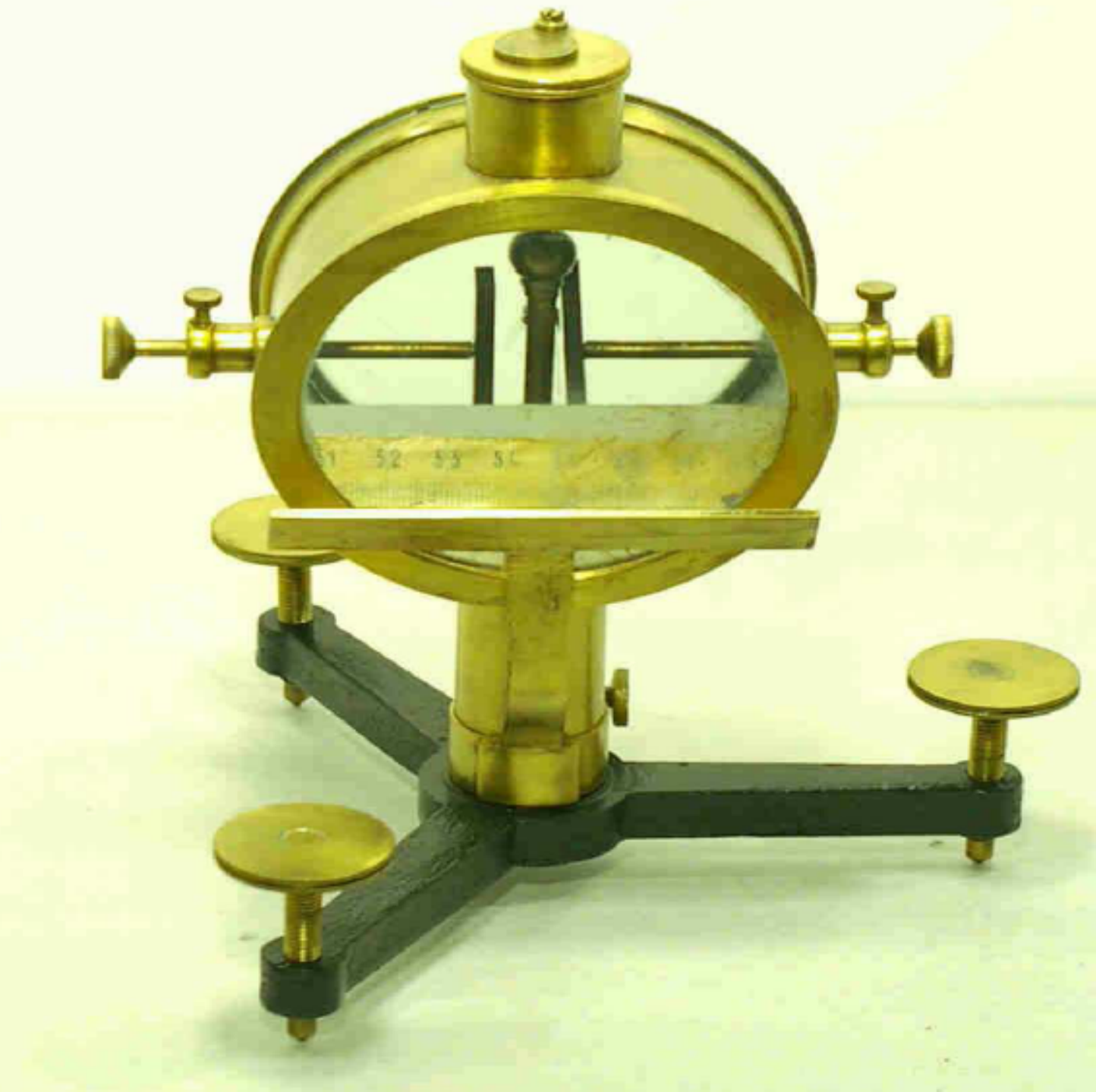} }
\end{center}
\caption{An Ebert electroscope of the beginning of 20th century. This instrument consists of an electroscope  through this a known flux of air could be pumped in.
Courtesy of Roberto Garra and of
the Gabinetto di Fisica dellÕ Istituto Calasanzio di Empoli.}
\label{fig:3}       
\end{figure}

During 1907-1911, he performed several detailed measurements on the conductivity of air, using an Ebert electroscope (Fig. \ref{fig:3}) to enhance the sensitivity of the measurement (he claimed he could reach a sensitivity of one third of volt). 

\begin{figure}
\begin{center}
\resizebox{0.75\columnwidth}{!}{ \includegraphics{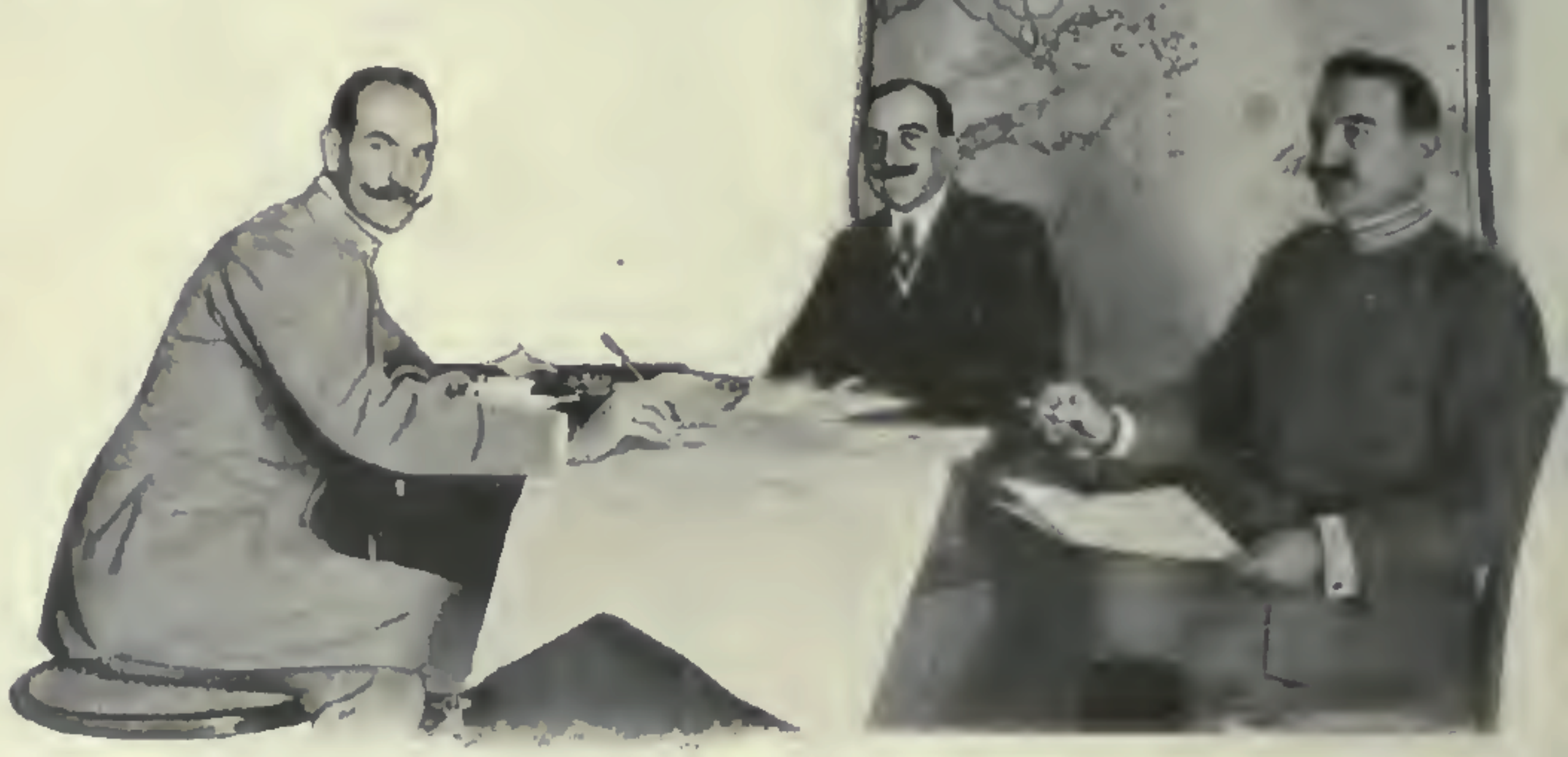} }
\end{center}
\caption{Gustavo Orsini (center), then  {\em{Capitano di Fregata,}} in 1913. Left is then Lt. Col. F.S. Grazioli, future hero of World War I  \cite{corsera}.}
\label{fig:orsini}       
\end{figure}

In 1907 and 1908 Pacini started a systematic measurement of the density of ions in Roma, in Sestola, in Livorno, on Mount Velino and in Forme, near Massa d'Albe, in the region of l'Aquila.
Additional measurements were  performed on the sea in the gulf of Genova, in front of the Accademia Navale di Livorno
 \cite{Pac1908a,Pac1908b} (on an Italian Navy  ship, the {\emph{cacciatorpediniere}} 
``Fulmine'', Fig. \ref{fig:fulmine}, commanded by the {\em{Capitano di Corvetta}} Gustavo Orsini, Fig. \ref{fig:orsini}). Also the President of the Central Bureau of Metereology and Geodynamics, professor Palazzo, participated to the first journey in the ``Fulmine'', in 1907; Palazzo documented the experiments related to atmospheric physics performed in that expedition in \cite{pala1}. This reference is also rich of photographic material about the expedition. Pacini tried to identify the sources of ionization, and in particular by studying the deactivation rate of a charged wire he could recognize the families of Radium, Thorium and Attinium;   such sources of radioactivity are present in the Earht's crust. 

In August and September 1908, Pacini started a systematic study of the time variation of the ionization in a chamber of Zinc 1.3 mm thick. He found strong variations (within a factor of five), depending on temperature, pressure, and humidity. He found also a daily cycle with two maxima; to explain such variations he made the hypothesis of a solar origin of part of the penetrating radiation, confirming  \cite{Ric1906}. 

As summarized by Cline~\cite{Cline} in 1910,
experiments at that time were mainly oriented
to measure the daily variations or seasonal variations of ionization. Cline cited the work by  Pacini~\cite{Pac1909} about the daily variations of the radiation measured on the
sea and at Sestola. Pacini's measurement was remarked in Cline's paper as a
first evidence of the atmosphere being the main responsible of the penetrating radiation,
excluding the Sun as the main contributor. 

The work by Pacini was also quoted by M. Curie in her {\em Trait\'e de radioactivit\'e} \cite{curietr}.

\begin{figure}
\begin{center}
\resizebox{0.9\columnwidth}{!}{ \includegraphics{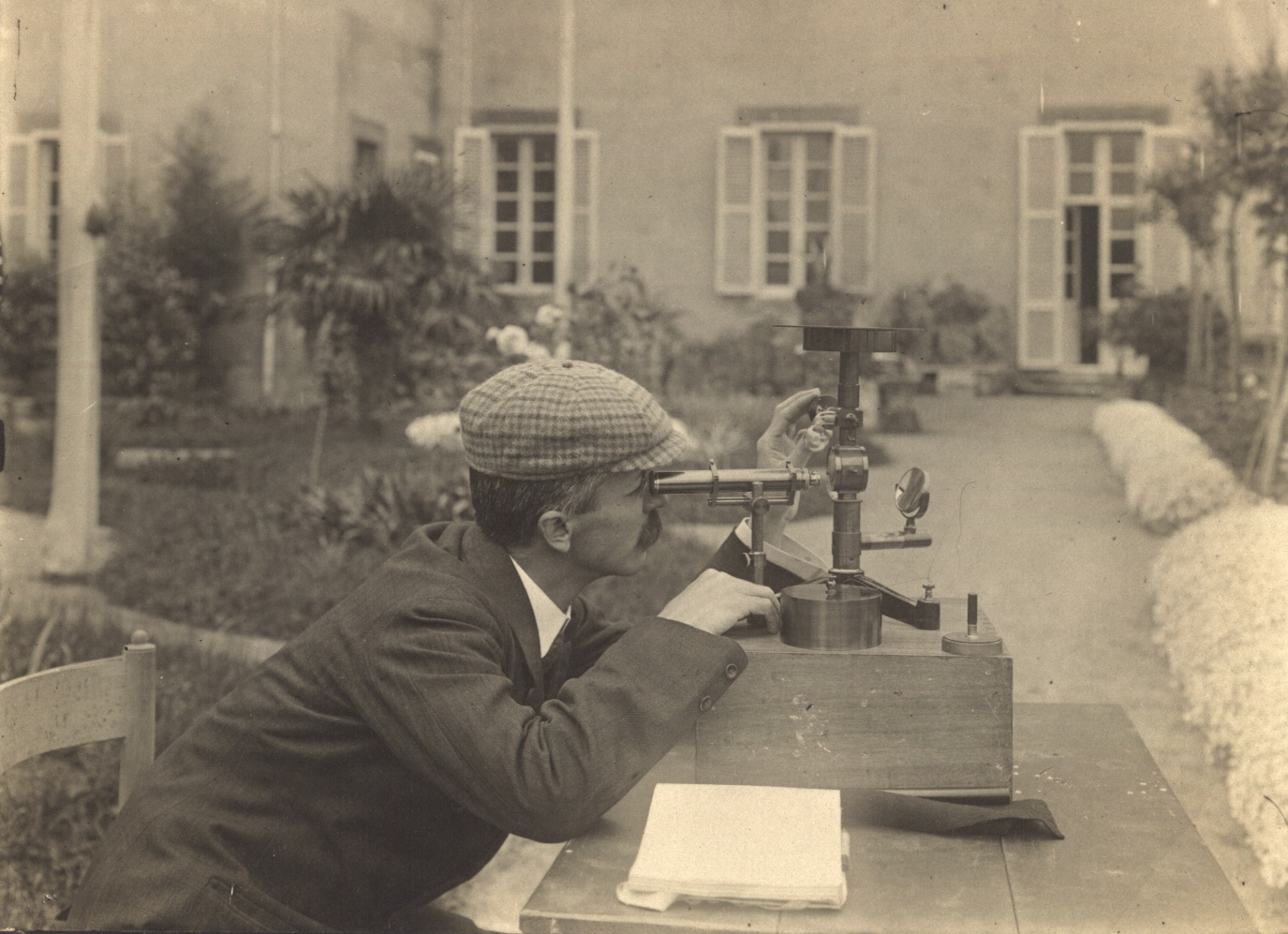} }
\end{center}
\caption{Pacini making a measurement in
May 1910. Courtesy of the Pacini family.}
\label{fig:pacinimisura}       
\end{figure}

Pacini started then an experimental program of systematic measurements of radiation(fig. \ref{fig:pacinimisura})
on the ground 
(at different elevations, including at sea level, and in different places to study local effects) and on the sea~\cite{Pac1910}. 
Those measurements  were aimed at checking 
whether the radioactivity within the Earth's crust was sufficient to explain the 
ionization effects (about 13 ions per second per cubic centimeter of air) that had been 
measured on the Earth's surface. 
In the paper \cite{Pac1910} Pacini found that the ionization on the sea surface, 300 m from the beach of Livorno, in front of the Naval Academy, was about two thirds of the ionization on the ground, 
thus supporting, at a variance with respect to the points of view of Wulf and Kurz, the idea that a non negligible part of the penetrating radiation  is independent of the emission from the soil. 

The definitive experiment is, however, the one performed in June 1911 \cite{Pac1912}, during 7 days of measurements in the deep sea in the Genova gulf, in front of the Naval Academy of Livorno. This measurement has an important place in the history of physics, since it pioneers the technique of underwater measurement of radiation.

With the apparatus at the surface of the sea 300 m from land, Pacini measured eight times during three hours
the discharge of the electroscope, obtaining an average loss of 12.6 V/hour, corresponding to
11.0 ions per second per cubic centimeter (with a RMS of 0.5 V/hour); with the apparatus at a 3 m depth in the 7 m deep sea, he measured 
as a result of seven trials an average loss of 10.3 V per hour, corresponding to 8.9 ions per second per cm$^3$
(with a RMS of 0.2 V/h). The difference (2.1 ions per second per cm$^3$) should be in the largest part (about 80\% in Pacini's estimate) attributed to a special radiation, independent of the radiation generated by the Earth's crust.

Consistent results were obtained during later measurements at
the Lake of Bracciano, where, at the same depth, the difference was of 2.2 ions per second.

The measurement underwater was 20\% lower than at the surface, consistent with absorption by water of a radiation coming from outside. ``With an absorption coefficient of 0.034 for water, it is easy to deduce from the known equation $I/I_0$ = exp(-d/$\lambda$), where d is the thickness of the matter crossed, that, in the conditions of my experiments, the activities of the sea-bed and of the surface were both negligible. É The explanation appears to be that, due to the absorbing power of water and the minimum amount of radioactive substances in the sea, absorption of radiation coming from the outside happens indeed, when the apparatus is immersed.'' Pacini concluded \cite{Pac1912}: ``[It] appears from the results of the work described in this Note that a sizable cause of ionization exists in the atmosphere, originating from penetrating radiation, independent of the direct action of radioactive substances in the soil." 

\begin{figure}
\begin{center}
\resizebox{0.6\columnwidth}{!}{ \includegraphics{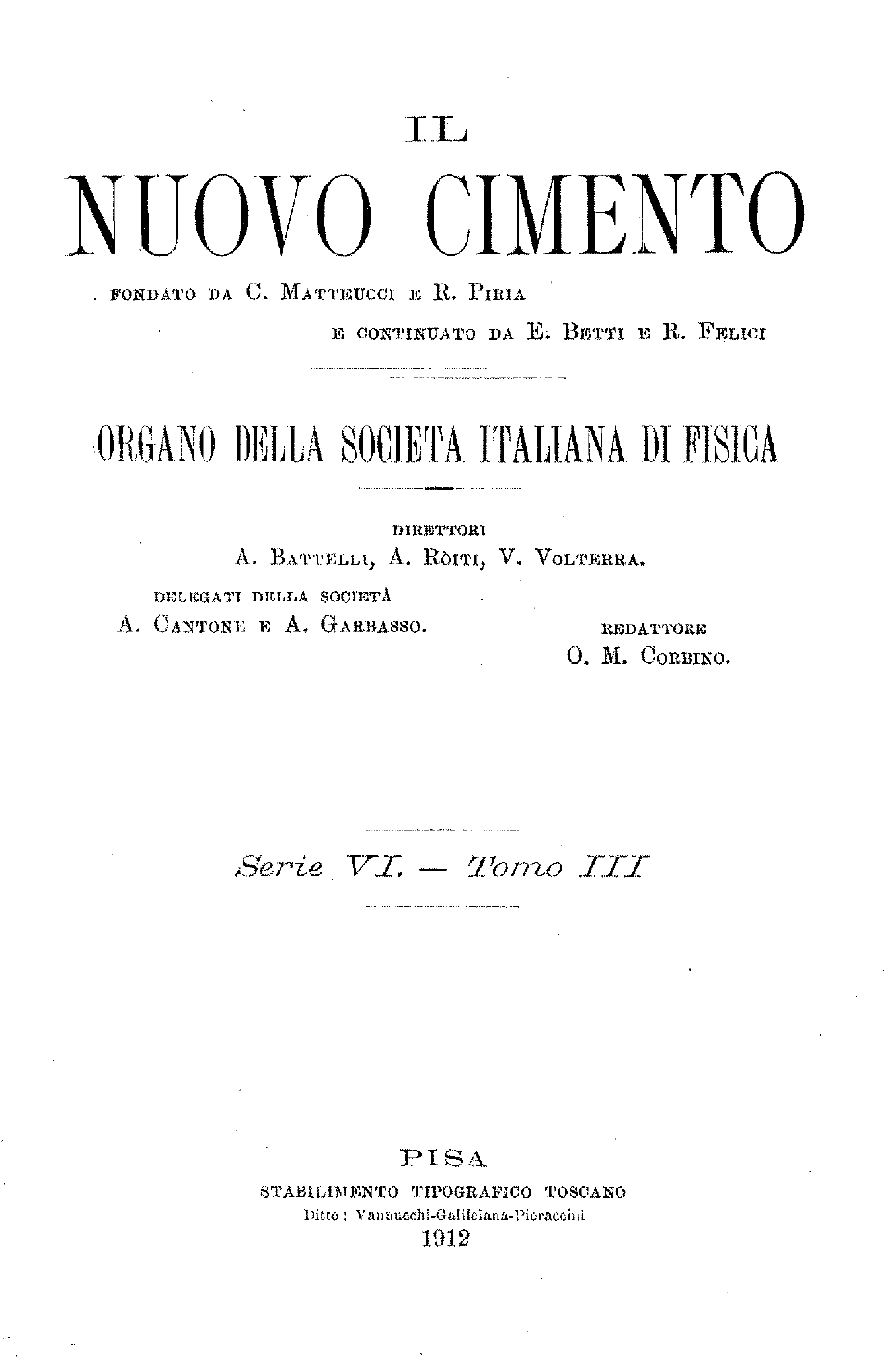} }
\end{center}
\caption{The cover page of the issue of ``Nuovo Cimento'' in which the article \cite{Pac1912} has been published.}
\label{fig:nccover}       
\end{figure}

Pacini's 
results, published in Nuovo Cimento in the early 1912 (Fig. \ref{fig:nccover}) marked the beginning of the underground/underwater technique for cosmic-ray studies 
(a technique that has been implemented so many times up to the present days).   

The technique by Pacini could not firmly exclude an atmospheric origin of radiation, but Pacini quotes Eve who claims that the contribution of radioactive substances in the air is negligible. 

We note that a measurement of  ionization  had been reported 1903 by McLennan and Burton \cite{MLB1903}, where
 the electroscope was placed in a container which in turn was closed in an iron tank, sometimes filled with 120 cm of water, sometimes shielded by metal layers. These authors observed a ionization reduction up to 40\%. Their conclusion was in line with radioactivity in the air from known radioactive sources: ``From these results it is evident that the ordinary air of a room is traversed by an exceedingly penetrating radiation such as that which Rutherford has shown to be emitted by thorium, radium and the excited radioactivity produced by thorium and radium'' \cite{MLB1903}.

As a curiosity, in 1910 Pacini had looked for a possible increase in radioactivity during a passage of the Halley's comet \cite{PaciniHalley}, and he found no 
evidence of a measurable effect from the comet itself.

In the beginning, Pacini was interested to ionization and radioactivity within a larger international research program related to the activities of the Central Bureau of Metereology and Geodynamics under the direction of professor Palazzo. Later on, after the clarification of the extraterrestrial origin of part of the ionizing radiation, Pacini somehow left the active research on this field (although he wrote review articles on the subject). The question was however still on, which part of this extraterrestrial radiation was coming from the Sun. After the works by Hess \cite{heclipse} and De Broglie \cite{debro}, also Palazzo himself \cite{pala2} made experiments on the subject. 

Finally, as pointed out in \cite{rob}, Italy had already in the beginning of the century aerostatic balloons capable to fly up to 5000 m; although Pacini had published in 1909 an article \cite{Pacballoon} on the perturbations produced by balloons on the Earth's electric field, he never organized a measurement of ionization.

\subsection{The correspondence between Pacini and Hess}

The first claims by Pacini \cite{Pac1909} came before any work by Hess. However in his
definitive paper \cite{Pac1912}, published a few months before the paper by Hess, Pacini was
evidently aware of the preliminary, and non conclusive (conclusive measurements by Hess came only in August 1912) results obtained by Hess
in 1911 as he did quote them correctly. Some excerpts from mail exchanges that occurred
between the two scientists in 1920, reported by \cite{Riz1934,bologna}, are very illuminating\footnote{In the fundamental paper in which he publishes for the first time a significant effect \cite{Hes1912}, Hess quotes the partial 1909 results by Pacini \cite{Pac1910}, but not his conclusive (statistically significant) results in 1911 \cite{Pac1912}. Unfortunately we do not know the times of submission of the two articles by Hess and Pacini; slowness of communication is a likely explanation. However, in the subsequent article by Hess, summarizing his measurements and the hypotheses on the origin of radiation \cite{Hes1913}, Pacini's works is not cited.}. 

On March 6, 1920, Pacini wrote to Hess: ``...I had the opportunity to study some of your papers about electrical-atmospherical phenomena that you submitted to the Principal Director of
the Central Bureau of Meteorology and Geodynamics [in Roma]. I was already aware of some of these works from summaries that had been reported to me during the war. [But] the paper \cite{Hess1919} entitled `Die Frage der durchdringenden Strahlung ausserterrestrischen Ursprunges' (`The problem of penetrating radiation of extraterrestrial origin') was unknown to
me. While I have to congratulate you for the clarity in which this important matter is
explained, I have to remark, unfortunately, that the Italian measurements and observations,
which take priority as far as the conclusions that you, Gockel and Kolh\"orster
draw, are missing; and I am so sorry about this, because in my own publications I
never forgot to mention and cite anyone...''\footnote{The original is in Italian: ``Egregio Signor Professore, ho potuto vedere alcune Sue pubblicazioni sui fenomeni elettro-atmosferici da
Lei spedite al direttore del R. Ufficio Centrale di Meteorologia e Geodinamica. Alcuni di questi lavori erano gi\`a noti a me nei riassunti che ne sono potuti pervenire durante la guerra. Mi \`e giunto quello nuovo intitolato `Die Frage der durchdrin. Strahlung ausserterrestrischen Ursprunges'. Mentre devo farle in proposito i miei complimenti per la chiarezza con cui espone in forma semplice lo stato della importante questione, mi duole che non siano stati citati affatto i lavori italiani su questo argomento, lavori a cui spetta senza dubbio la priorit\`a, per quanto si riferisce alla previsione delle importantissime conclusioni a cui sono successivamente pervenuti il Gockel, Ella stessa, signor Hess, ed il Kolh\"orster; e tanto pi\`u me ne duole, in quanto, nelle mie pubblicazioni, io non ho mai dimenticato di citare
chi di dovere. Con perfetta osservanza. Prof. Dr. Domenico Pacini''.}.

The answer by Hess, dated March 17, 1920, was: ``Dear Mr. Professor, your very
valuable letter dated March 6 was to me particularly precious because it gave me the
opportunity to re-establish our links that unfortunately were severed during the war.
I could have contacted you before, but unfortunately I did not know your address. My
short paper `Die Frage der durchdringenden Strahlung ausserterrestrischen Ursprunges'
is a report of a public conference, and therefore has no claim of completeness. Since
it reported the first balloon measurements, I did not provide an in-depth explanation
of your sea measurements, which are well known to me. Therefore please excuse me
for my unkind omission, that was truly far from my aim ...''\footnote{The original \cite{bologna} is in German: ``... und macht, hinsichtlich der Literaturangaben, keinerlei Anspruch auf Vollst\"andigkeit. Da es in erster Linie auf die Messungen im Ballon ankam, bin ich auf Ihre mir wohlbekannten Messungen \"uber die Meere nicht speziell eingegangen. Ich bitte diese Unterlassung g\"utigst zu entschuldigen; mir ist dabei jede Absicht ferne gelegen.''}. 

On April 12, 1920,
Pacini in turn replied to Hess: ``... [W]hat you say about the measurements on the
penetrating radiation performed on balloon is correct; however the paper `Die Frage
der durchdringenden Strahlung ausserterrestrischen Ursprunges' lingers quite a bit on
measurements of the attenuation of this radiation made before your balloon flights,
and several authors are cited whereas I do not see any reference to my relevant
measurements (on the same matter) performed underwater in the sea and in the
Bracciano Lake, that led me to the same conclusions that the balloon flights have
later confirmed.''\footnote{The original is in Italian: ``[S]ta benissimo quanto Ella mi dice circa le misure di radiazione penetrante eseguite su pallone aerostatico,
tuttavia nella Sua pubblicazione `Die Frage der durchdringenden Strahlung ausserterrestrischen Ursprunges' si parla a lungo
delle misure fatte per stabilire l'assorbimento di queste radiazioni e si citano vari autori, mentre non vedo citati i
miei lavori in proposito, eseguiti in seno alle acque del mare e in seno alle acque del lago di Bracciano, lavori dai quali
potetti dedurre il valore della radiazione penetrante nell'aria e trarre conclusioni che le esperienze a bordo di un pallone
aerostatico hanno poi
confermato.''}

Finally, on May 20, 1920, Hess replied to Pacini: ``Coming back to your publication in `Nuovo Cimento', (6) 3 Vol. 93, February 1912, I am ready to acknowledge that certainly you had the priority in expressing the statement, that a non terrestrial radiation of 2 ions/cm$^3$ per second at sea level is present.  However, the demonstration of
the existence of a new
source of penetrating radiation
from above
came from my balloon ascent to a height of 5000 meters on August 7 1912, in which I have discovered a huge increase in radiation above 3000 meters.''\footnote{The original is in German: ``...Zur\"uckkommend auf Ihre Publikation in "Nuovo Cimento", (6) 3 Bd 93, 1912 Februar, erkenne ich gerne an, dass Ihnen hinsichtlich der Aeusserung der Vermutung, dass eine nicht von der Erde ausgehende Strahlung von ca. 2 Jonen/cm$^3$/s (in Meeresh\"ohe) vorhanden sei, unbedingt Priorit\"at geb\"uhrt. Die Gewissheit \"uber das Vorhandensein einer von oben kommenden neuen Quelle der durchdringenden Strahlung hat aber doch erst meine Ballonfahrt auf 5.000 m vom 7. August 1912 gebracht, bei welcher ich zuerst unzweifelhaft eine enorme Vergr\"osserung der Strahlung \"uber 3.000 m gefunden habe. ...''}

The Hess-Pacini correspondence, 9 years after Pacini's work and 8 years after Hess' 1912 balloon flight, shows how difficult communication was at the time. Also language difficulties may have contributed: Pacini publishing mostly in Italian and Hess in German. Finally, due to the lack of academic freedom, Pacini could not attend the conference in Wien in September 1913 (85th Versammlung Deutscher Naturforscher und \"Artze) in which  the new results 
on the origin of the penetrating radiation were discussed \cite{libropd}.




\section{Cosmic rays: the recognition by the scientific community}

Hess is today remembered as the discoverer of cosmic rays for which he was awarded the 1936 Nobel Prize in physics, nominated by Compton. In his nomination Compton had written: ``The time has now arrived, it seems to me, when we can say that the so-called cosmic rays definitely have their origin at such remote distances from the Earth that they may properly be called cosmic, and that the use of the rays has by now led to results of such importance that they may be considered a discovery of the first magnitude. [...] It is, I believe, correct to say that Hess was the first to establish the increase of the ionization observed in electroscopes with increasing altitude; and he was certainly the first to ascribe with confidence this increased ionization to radiation coming from outside the Earth''. Why so late a recognition? Compton writes: ``Before it was appropriate to award the Nobel Prize for the discovery of these rays, it was necessary to await more positive evidence regarding their unique characteristics and importance in various fields of physics" \cite{noi2}.   The Nobel prize to Hess was shared with
C.D. Anderson for the discovery of the positron.

Hess' discovery was based on contributions of many other scientists. 
It seems to us that in particular the important contribution by Pacini has been forgotten in the history of science. Pacini reached important conclusions on the origin of the ``penetrating radiation'' one year before Hess; the technique used by Pacini, anyway, could not firmly disprove a possible atmospheric origin of
the background radiation.  Besides the fact that Pacini died in 1934, two years before the Nobel to Hess, one has to note that he had never been nominated for the Nobel prize - although his work had been cited in the final report by the Nobel prize Committee to the Royal Academy of Sweden. Such report quotes Gockel's conclusion that the results of his balloon measurements, in agreement with measurements by Pacini, show that a not insignificant part of the radiation is independent of direct action of substances in the crust of the Earth;  notes however that Hess' careful work includes an accurate measurement of the absorption of gamma rays, confirming the results of Eve, and several balloon ascents in 1911 and 1912, and shows for the first time that a very penetrating radiation is incident on the atmosphere from the outside \cite{noi2}.

Scientific research is today characterized by openness and rapid communication of results. This was not the case when cosmic rays were discovered \cite{noi2}. Communication was slow, there were language barriers combined with nationalism and there were important effects of World War I. Several other causes might have contributed to the lack of credit given in present days to Pacini's work \cite{rob,stra1,stra2}, last but not least rivalries and controversies between Europe and the US \cite{noi2,craw,demaria}.

\begin{figure}
\begin{center}
\resizebox{\columnwidth}{!}{%
\includegraphics{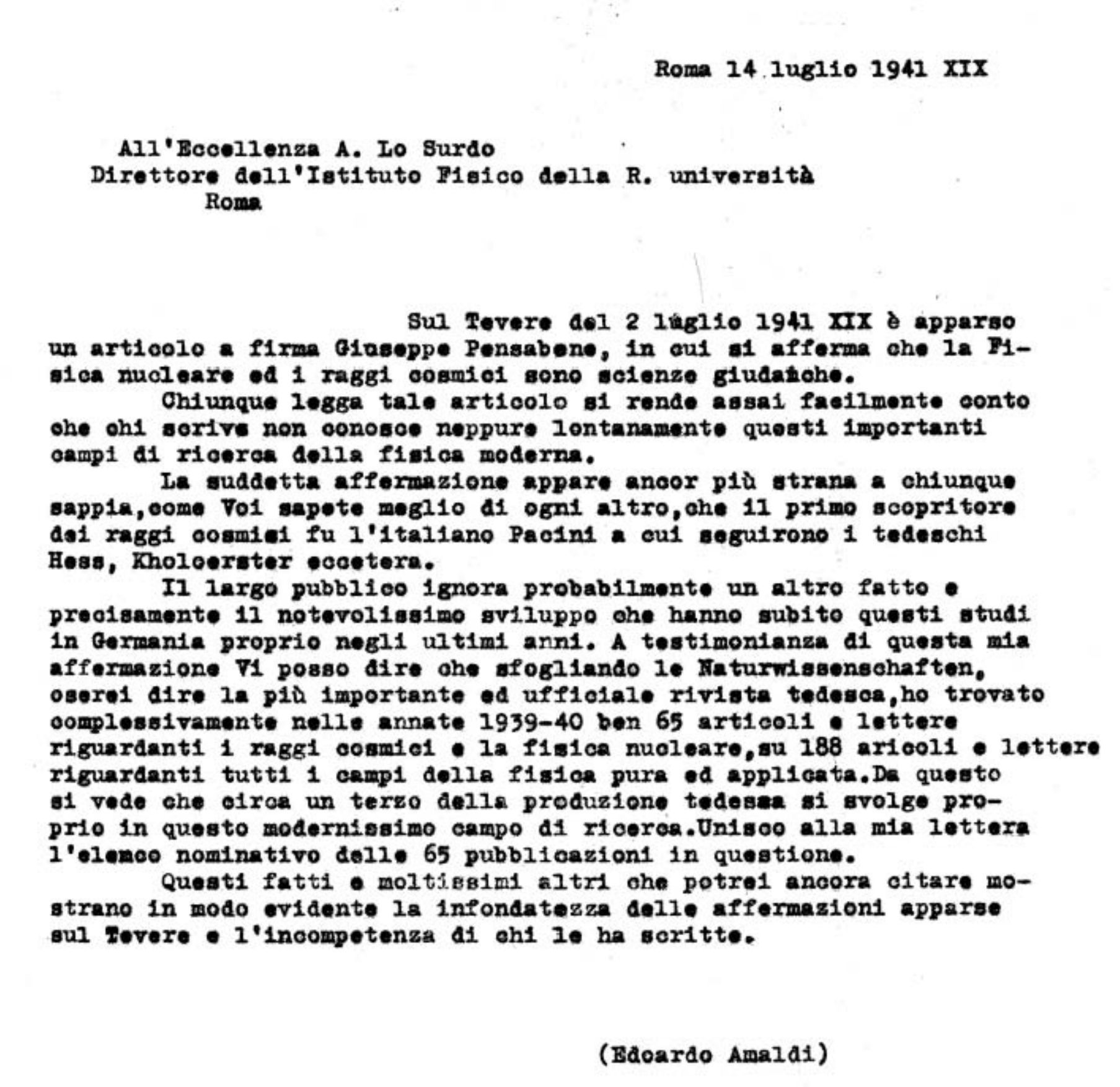} }
\end{center}
\caption{Letter by E. Amaldi to Lo Surdo. The letter belongs to the Amaldi Archive (scatola 212, fasc. 1) at the ``La Sapienza'' University of Roma (courtesy of U. Amaldi and G. Battimelli). A translation of the text follows. 
{\em ``In the newspaper `Il Tevere', July 2, 1941, an article signed by
Giuseppe Pensabene was published, saying that nuclear  and cosmic ray
physics are Judaic sciences.
Readers can easily understand that the author of that article does not
know at any level such important research fields in modern physics.
That opinion appears even stranger to anyone who knows, and you know
better than any other, that the first discoverer of cosmic rays was
the Italian Pacini, followed by the Germans [sic!] Hess,
Kholhoerster, etc.
The public ignores probably another fact: such fields had a large
development in Germany in the recent years. This is documented by the
fact that on Naturwissenshaften, probably the most important and
official German journal, I found, in the years 1939-40, 65 articles on
cosmic rays and nuclear physics, out of a total of 188 articles on the
whole field of physics, pure and applied. One can conclude that about one third of the
German production is related to such extremely modern subjects. I enclose the list of
these 65 articles.
Such facts, and many more that I could quote, show evidently that what
is written in `Il Tevere' has no foundation whatever, and illustrate the
incompetence of the author.''}}
\label{fig:amaldilet}       
\end{figure}

Edoardo Amaldi\footnote{Edoardo Amaldi (Carpaneto Piacentino, 1908 - Roma, 1989) worked closely with Enrico Fermi (he had been  one of the ``boys from Via Panisperna'') until 1938, when Fermi 
was forced to leave Italy. Professor of experimental physics in Roma for more than 40 years, he was co-founder of CERN, of which he was Secretary-General (this was the name of what is called today Director-General) during the years 1952-1954, of ESA, and of the Italian National Institute for Nuclear Physics (INFN), of which he was President during the years 1960-1965.
His fields of interest were elementary particle physics, nuclear physics, and the physics of cosmic radiation, with a particular attention to the study of gravitational waves.} 
 had no doubt that Domenico Pacini was indeed the discoverer 
of cosmic rays as stated in a letter (Fig. \ref{fig:amaldilet})  that he wrote on July 14, 1941 to the director of the Physics Institute of Roma, Antonino 
Lo Surdo\footnote{Antonino Lo Surdo (Siracusa, 1880 - Roma,  1949) was professor of physics in the University of Roma since 1919; he became director of the Physics Institute in 1937. Founder of the National Institute of Geophysics, he was expelled after World War II 
 from the Accademia dei Lincei, one of the most prestigious scientific academies in Italy, because of 
claims of collaboration with Mussolini's regime, and later 
reintegrated \cite{forestamartin}.}.

E. Amaldi's letter was motivated by an article that had appeared on July 2, 1941 in a local newspaper, ``Il Tevere'', in which it was stated that nuclear physics and cosmic ray physics were ``Jewish sciences''. Although ``Il Tevere'' was not the official journal of the Fascist Party,
it had anyway a large political influence, as it was known that its content was dictated by Benito Mussolini. E. Amaldi writes that such a statement,
potentially dangerous for fundamental science in Italy, ``...appears so strange to anyone who
knows, as you certainly do, that the Italian Domenico Pacini [(a non-Jew)] was the discoverer of
the cosmic rays ...'' 

Is it appropriate to say, as Edoardo Amaldi did, that  Pacini was the discoverer of the cosmic rays and was followed by
 Hess, Kolh\"orster, etc.? A great discovery is in general the
result of joint efforts by many people. It is certainly true that Pacini wrote,
as early as 1909, that the action of active substances in the soil was
not sufficient to explain the observed properties of the penetrating radiation;
and that he was the first to draw this conclusion based on experimental data, providing a conclusive demonstration in 1911 (one year before Hess,
who knew about Pacini's results when he published his famous article).
 It is also true that Pacini's measurements in June 1911 were a breakthrough, since the technique of measurement below the sea surface,
 using water as a shield, started thanks to his experiment. One should however recognize that Pacini's experiment, which proved that the radiation was coming in large part from outside the Earth's crust, could not firmly exclude the atmosphere as a possible source. Professor H. Pleijel, Chairman of the Nobel Committee for Physics of the Royal Swedish Academy of Sciences, said in his speech~\cite{nobelsp} at the 
 Nobel award ceremony on December 10, 1936: ``[A] search was made throughout nature for radioactive substances [by several scientists]: in the crust of the Earth, in the seas, and in the atmosphere; and the instrument just mentioned -the electroscope- was applied. Radioactive rays were found everywhere, whether investigations were made into the waters of deep lakes, or into high mountains. ... Although no definite results were gained from these investigations, they did show that the omnipresent radiation could not be attributed to radiation of radioactive substances in the Earth's crust.'' And then, ``The mystery of the origin of this radiation remained [however] unsolved until Prof. Hess made it his problem. ... With superb experimental skill Hess perfected the instrumental equipment used and eliminated its sources of error. With these preparations completed, Hess made a number of balloon ascents... From these investigations Hess drew the conclusion that there exists an extremely penetrating radiation coming from space which enters the Earth's atmosphere.'' 
 
In any case, a whole
community of researchers was involved in that field. Pacini certainly made a break-through by
introducing the technique of underwater measurement and by finding a significant decrease of the radiation with respect to the surface, which could exclude the Earth as the only source of radiation; 
unfortunately, he could not participate properly to the ongoing debate,
and he could not push his results with energy. Pacini's work
was carried out in difficult conditions because of lack of resources available
to him, because of lack of scientific freedom during the crucial years when he
was working at the Central Bureau of Meteorology and Geodynamics, and
because of the substantial indifference his work was met with by the
Italian academic world - that, {\em{per se,}} made Nobel Prize nominations difficult. 

The work behind the discovery of cosmic rays, a milestone in science, comprised scientists in Europe
and the US and took place during a period characterized by lack of communication and by nationalism caused primarily by the World War I.
In the work that culminated with high altitude balloon flights,  important contributions have been forgotten and in particular those of Pacini. Besides the personal stories related to the accidents of the lives of individuals, several historical and political facts contributed to the lack of references to the work of Pacini in the history of cosmic rays, and should serve as a lesson for the
scientific politics of the future.

%

\section*{Acknowledgements}

The University of Bari,  Professor A.~Garuccio in particular, 
supported the research of documents related to Domenico Pacini;  the Dipartimento 
Interateneo di Fisica of Bari 
organized the Domenico Pacini memorial day that 
was held in Bari on April 17, 2007.

Professors F.~Guerra and N.~Robotti uncovered the letter by E. Amaldi to ``Il Tevere'' and the dossier of the applications by Pacini for the ``libera docenza'' and for the professorship in the Amaldi Archive at Roma's ``La Sapienza'' University and in the Biblioteca Nazionale, respectively; 
they also provided an in-depth analysis of the scientific path leading Pacini to the hypothesis of extraterrestrial radiation  \cite{rob}. Several aspects were clarified thanks to discussion with Professor Guerra. R. Garra uncovered material in ``Collegio Romano''. I thank Professors U. Amaldi and G. Battimelli for allowing  to reproduce the letter by E. Amaldi.

Comandante E.~Bagnasco and A.~Lombardi from Associazione Culturale Italia, C.~D'A\-damo and H. von Zeschau from Associazione Regia Marina Italiana, Ten. Col. M. Piemontese from Guardia di Finanza and Dr. M.L. Rischitelli helped in retrieving historical pictures and information on the ships by the Italian Navy.

The AMA - Servizi Cimiteriali di Roma and Don Mario Del Turco from Forme kindly provided me with information on the remains of Pacini. Don Mario and the clerk Ada Libertini from the City of Massa d'Albe gave me information on the Pacini family.

Large part of the original material comes from the Pacini family, and in particular from Dr. Lia Santoponte, Dr. Giovanna Minardi Zincone, Professor Sergio Zincone, Benedetto Valente, Professor Cesare Bacci. Transcriptions and translations from German were done by Dr. B. Steinke from MPI Munich.
 
Professors B. De Lotto, L.~Guerriero, E.~Menichetti, R.~Paoletti, M.~Persic  and Dr. P.~Lipari, provided help, support, suggestions, and comments. Besides suggestions and
comments, Professor P.~Spinelli provided unpublished material and invited me to the Pacini memorial day. Several aspects of the research on cosmic rays became clearer to me thanks to discussions with Professors H. Rechenberg and A.A. Watson.

Professor P. Carlson uncovered relevant material related to the Nobel prize to Hess, and coauthored Ref. \cite{noi2}; discussions with him clarified 
in addition several important points, and were enlightening related to the problem of nationalism and internationalism in science. 
Professor K. Grandin at the Center for History of Science at the Royal Swedish
Academy of Sciences provided help and support for the study of the Academy's archives. 

Dr. M. De Maria and L. Marafatto did most of the work for the translation of the articles in Appendix 2 and Appendix 1, respectively. Dr. S. De Angelis from Williams Language Solutions and Dr. S. Kim
from Blue Pencil Science helped in the translation and editing of the article in Appendix 3.

Professors L.~Cifarelli and  E. De Sanctis encouraged me since 2007 in the study of Pacini's work. 
Professor A. Bettini invited me to document the work by Pacini for the Rivista del Nuovo Cimento of the Societ\`a Italiana di Fisica in the centennial of Pacini's discovery; he also gave me 
precious suggestions and clarifications on the studies related to light scattering in the atmosphere.  
The editorial office of SIF in Bologna, Dr. Marcella Missiroli in particular, helped in the final editing, and solved copyright questions for the reprinted articles.


I would like to thank the Max Planck Institute in Munich, and Professor M. Teshima, Dr. R. Mirzoyan and the MAGIC MPI group, for the kind hospitality in Germany.

A special acknowledgement goes finally to Professors  N. Giglietto and S. Stramaglia from Bari, who started the rediscovery of the work from Pacini, uncovered a lot of material, and transmitted me their enthusiasm for giving an unbiased account of this uncredited contribution to the discovery of cosmic rays.


%


\clearpage

\appendix

\noindent{\textsc{\large{\bf{Appendix}}}}\vskip 2 truemm

\noindent{\bf{\large{Three selected articles by Pacini, translated and commented}}}
\addcontentsline{toc}{section}{Appendix: Three selected articles by Pacini, translated and commented}


\vskip 0.5 truecm

We reprint below the translation of three selected papers by Pacini, written between 1908 and 1911. 

They have been chosen since, in the opinion of the Author of this review, they best represent the original contribution by Pacini to the pioneering phase of the interpretation of the nature and origin of cosmic rays; however, other pieces of experimental work by Pacini are as relevant or more relevant for the study of atmospheric ionization. We hope that all Pacini's articles will be 
translated and published in a near future; an almost complete selection of the original papers related to cosmic rays can be found in \cite{deacoll}.

The paper in Appendix B comes from the French translation of an original article published by Pacini in Italian; the translation into French was not done by Pacini himself, but the Author of this review thinks that this version had to be taken as original since it comes from a late revision of the original article by Pacini himself.

The three articles show the evolution of the interpretation of Pacini towards his final conclusion, adding piece by piece missing experimental evidence
by means of the design and execution of dedicated measurements. This is a paradigm of a successful scientific research.
\begin{itemize}
\item[A.] In the first paper Pacini measures daily periodical variations and fluctuations of ionization with time, and concludes that ``[i]n the hypothesis that the origin of penetrating radiations is in the soil [...] it is not possible to explain the results obtained.''
\item[B.] In the second paper Pacini compares the air ionization on the sea's surface and on ground; he finds that the radiation is slightly smaller at the sea's surface, and concludes that ``results seem to indicate that {\em
a substantial part of the penetrating radiation in the air, especially the one that is subject to significant fluctuations, has an origin independent of the direct action of active substances in the upper layers the Earth's crust.''\footnote{In italics in the original.}}
\item[C.] In the third paper Pacini adds a decisive experimental evidence, measuring a significant decrease of the ionization rate underwater with respect to the sea's surface, and concludes that 
{\em a sizable cause of ionization exists in the atmosphere, originating from penetrating radiation, 
independent of the 
direct action of radioactive substances in the soil.''\footnote{In italics in the original.}}
\end{itemize}

\setcounter{footnote}{0}

\clearpage





\section{\BY{D. Pacini} {\em ``On penetrating radiations'',} \IN{Rend. Acc. Lincei} {18}{1909} {123}, translated and commented by L. Marafatto and A. De Angelis}

\vskip 1cm

\begin{center}{\bf{ {\em PHYSICS}  -- {ON PENETRATING RADIATIONS}}}

{{Note by D. Pacini, presented  by our Member P. Blaserna}}\end{center}

\vskip 1cm

The question of the origin of penetrating radiations capable of ionizing the gas enclosed in a metal container, and coming from external agents, is presently subject of discussion. It is useful to measure this effect in different places, in order to study the intensity of such radiations, their variation with time, and to investigate possible correlations with other phenomena known in atmospheric physics.

A Zinc plate\footnote{In accordance with the latest experimental results, and in particular with those obtained by Mc. Lennan (Phil. Mag., December 1907) on the conductivity of air enclosed in metallic containers, I used Zinc. Samples of Zinc with little contamination from radioactive substances are relatively easy to obtain.} deprived of the surface layer, carefully wiped and polished, was used to build the cylindrical container used for the experiments that I made in open air in the countryside. This container was 1.3 mm thick, 59.5 cm high, 12 cm in radius. 
The internal electrode was a copper cylinder which was connected  through an insulator to a small Aluminum electroscope put outside. A simple device allowed to get a potential difference varying from 300 V to 450 V, setting the leaf of the electroscope at an appropriate angle; so in this range it was possible to get the saturation current. The electrometer was equipped with a microscope which allowed to measure the drop of the potential in the range with an accuracy of 1/3 of volt. The capacity of the system, measured with a Harms condenser, was of about 19 cm. 
The Zinc cylinder had two faucets for air exchange. Some days before beginning the measurements\footnote{Wood and Campbell in Phil. Mag. 1907, 265, find that for a gas enclosed in a Zinc container, the increase of ionization with time is almost negligible.} air was introduced in the container through a cotton filter. The electrometer was protected by a black iron box, with a glass window in front. The device, protected on the upper side by a waterproof curtain, was assembled at Sestola's meteorological observatory, on an insulated hill at about 1090 m a.s.l. on Modena's Apennines. Measurements were taken in August and September 1908. 
Since the experiment was performed in open air, taking measures every hour, the observation series are complete for the whole 24 hours in days without atmospheric disturbances. At the same time and also hour by hour, but only during daytime, Ebert measurements of free air ionization\footnote{I want to thank dr. Carlo Zanini for his kind help in the series of Ebert measurements and in the meteorological observations.} were taken. Meteorological observations were always made together with every reading of the electrometer.

The own dispersion of the instrument, being the insulation good, was very small during the interval between measurements, and was found to be constant every time it was measured, since the electrometer was protected by a metallic box. Assuming $3.4 \times 10^{-10}$  u.e.a. as the value of the ion charge, one can easily obtain from the rate of decrease of the potential in a given time interval, the value usually referred as $q,$ {\em{i.e.,}} the number of ions produced in one second in every cm$^{3}$ of air, in the container.

On the first row of the following table the values obtained from the average of the values  of $q$ at the same hour in different observation days are reported; these correspond to the series of experiments made  every day in which the sky was predominantly clear and the air was calm during about 2 months. In the next lines one can see similar averages, related to the quantities of electricity $I_+$ and $I_-$  expressed in u.e.a.; to the value of unipolarity  $Q={I_+}/{I_-}$ and to data related to the variation of meteorological elements. The plot is the graphical translation of the measurements.

\begin{center}{\em{Average value of the ionization in the closed container in a single hour of the day.}}\end{center}

\begin{center}
\resizebox{\columnwidth}{!}{\includegraphics{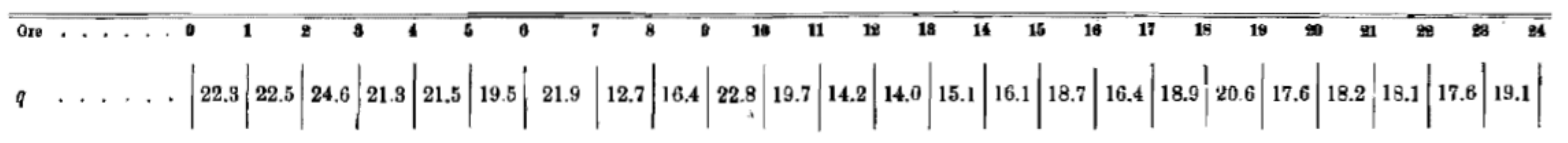} }
\end{center}

\begin{center}{\em{Average value, at a given time, of the quantities: ($I_+$,$I_-$, $Q={I_+}/{I_-}$, Pressure, Relative humidity, Temperature). (1) The values for 
$I_+,$ $I_-$ and $Q$ observed at 7 a.m., which due to the relatively small number of observations could be inaccurate, are marked by an asterisk. In the Figure, tracts which correspond to these data are drawn as a dotted line.}}\end{center}
 
\begin{center}
\resizebox{\columnwidth}{!}{\includegraphics{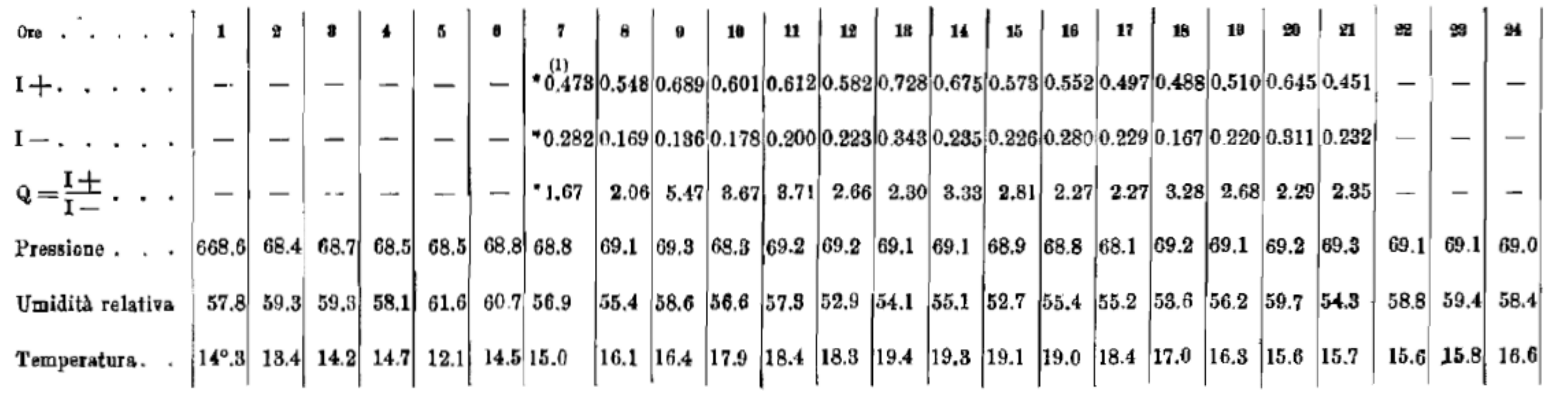} }
\end{center}

\vskip 0.3cm
\noindent {\bf{Experimental results:}}
\vskip 0.3cm

{\itshape The minimum value obtained in Sestola for the air ionization inside a Zinc container was of 6.3 ions per cm$^{3}$ per second}, and it was observed between 7 and 8 a.m.,  {\em{i.e.,}} when ionization outside was always minimum; and it is worth noticing that sometimes we got values of $q$ = 8, $q$ = 9, and many times values of 10 and 12, all quite low. The maximum value of 30 ions was obtained between 11 p.m. and 3 a.m.,  {\em{i.e.,}} corresponding to the maximum of the ionization of the air outside.

\begin{center}
\resizebox{\columnwidth}{!}{\includegraphics{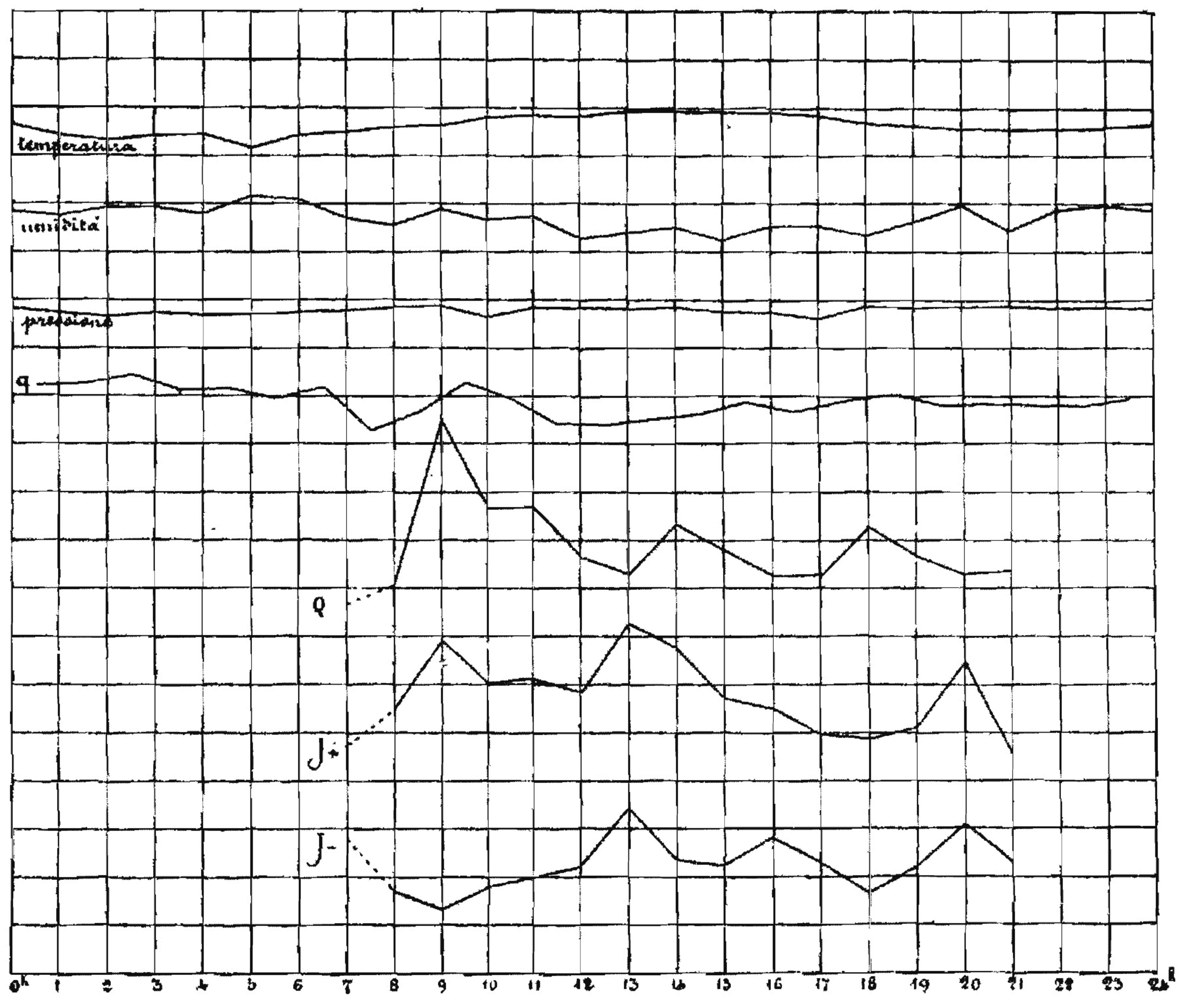} }
\end{center}

It is important to observe that the value $q$ = 6.0 for ionization in a closed vessel, measured at standard pressure and temperature, without any further absorbing shields, is the lowest so far obtained by Mc. Lennan\footnote{J.C. Mc. Lennan, {\em{Phys. Rev.}} {\bf 526}, 1908.}, who however made his measurements on the ice of the Toronto bay. The value of 6.3 is the minimum observed on mainland above rocky places. The trend of the curve is fundamentally different from what obtained by Strong\footnote{W.W. Strong, {\em{Phys. Zeit.}} {\bf 118}, 1908.} who made experiments similar to the ones by the writer in open country (in summer 1907 in Pennsylvania in Cumberland valley, 7 miles away from Harrisburg). During these first researches on the subject, Strong made experiments in the air contained in a small Aluminum electroscope, and found that internal ionization shows a maximum during the day near noon and a minimum during the night;  this result is completely different from that of other experimenters and also different, as we will see now, from results obtained later by Strong himself.

{\itshape Ionization of air contained in a metallic container exposed to open air in the countryside shows  a double daytime peak in Sestola too}: there is a first maximum after midnight, which becomes more noticeable between 2 and 3 a.m.; a second one between 9 and 10 a.m.; there is a minimum between 7 and 8 a.m., another between 12 a.m. and 1 p.m..
Related to the trend of the penetrating radiation, here are the results obtained up to now in different places.


\noindent\begin{minipage}{\textwidth}
\begin{small}
\noindent
\begin{tabular}{l|c|c|c|c|c|c}
 & {Peters-}  & 
 {Wien} & 
 {Cambridge}  & 
 {Washing-}   & 
 {Baltimora}  & 
 {Sestola} \\
 & {burg}\footnotemark[6]  & 
\footnotemark[7] & 
 \footnotemark[8]  & 
ton\footnotemark[9]  & 
 \footnotemark[10]  & 
  \\
Max. &  & morn./eve. & 8$^h$-10$^h$; 22$^h$-1$^h$ & 10$^h$ ; 23 $^h$ & 9$^h$ ; 22$^h$ & 2$^h$-3$^h$ ; 9$^h$-10$^h$ \\
Min. & $\approx$ ~15$^h$ & noon & 4$^h$ ; 14$^h$ & 7$^h$ ; 18$^h$ & 7$^h$ ; 18$^h$ &  7$^h$-8$^h$ ; 12$^h$-13$^h$ \\
\end{tabular}      
\end{small}
\end{minipage}

\vskip 0.5cm

\setcounter{footnote}{5}

\footnotetext[6]{J.J. Borgmann, {\em Science Abstracts,} 1905, n. 1580.}
\footnotetext[7]{H. Mache and T. Rimmer, {\em Phis. Zeit.} {617}, 1906.} 
\footnotetext[8]{Wood and Campbell, {\em Phil. Mag.} 265, 1907.}
\footnotetext[9]{T.F. Mc. Keon, {\em Phys. Rev.} {\bf 25}, 399, year 1907.}
\footnotetext[10]{W.W. Strong, {\em Phys. Rev.}  {\bf 27}, 39, year 1908. Experiments which brought to the determination of time-interval were made by the author in the tower of the physics laboratory of the Johns Hopkins University. A curve obtained at Mechanicsburg Pa. on December 26$^{th}$, 1907, shows the same trend. The author does not report if he obtained such result it in the open air or in a closed setting.} 

\setcounter{footnote}{10}

The ionization in free air displays its maximum 
in correspondence with the minimum of penetrating radiation at about 1 p.m.; the unipolarity value $Q$ shows oscillations nearly matching the ionization of the Zinc vessel.
The curves obtained from the average values of pressure, relative humidity and temperature, show that these meteorological elements undergo relatively small variations during the observation days, and that there is no clear correlation between their variation and the ionization phenomenon in the metallic container.

\vskip 0.3cm
\noindent {\bf{Discussion of the results.}}
\vskip 0.3cm
 
 In the work cited, Mc. Lennan observes that, making experiments on land, one gets values for $q$ larger than those obtained on the layer of water and ice in the Toronto bay and on the Ontario lake, because, according to the Author's experiences, water and ice should absorb most of the radiation coming from the submerged soil. In addition he notes that while on a 26 meters high tower  value $q = 11.4$ is measured, one gets $q = 15.3$
in a room in the basement of the building itself, which, being recently built,  is supposed to be free from radioactive impurities. From these facts the Author gathers an argument against the hypothesis which attributes to air the origin of penetrating radiations.

Many times during the various measurements in Sestola I obtained values for $q$ lower than 10, down  to the minimum of 6.3. Admitting with Mc. Lennan that the small values he got for $q$ above water are of same order of magnitude of the effects that can be produced by residual radioactive impurities in metals, one must deduce that it is sometimes possible to observe also on mainland and rocks an almost null effect from penetrating radiations coming from outside, and from the secondary radiations due to the interaction of penetrating radiation on the container walls, on the ionization of the air contained in a closed vessel. This fact, also admitting that the limestone soil above which I made my measurements contains a relatively small amount of active substances,  suggests some considerations about the question if the origin of penetrating radiation is or is not in the soil in consideration of the amplitude and the daytime trend of the oscillation of the ionization values.

In the hypothesis that the origin of penetrating radiations is in the soil, since one must admit that they are emitted  at an almost constant rate (at least when the soil is not covered by recent precipitations), it is not possible to explain the results obtained up to now, and in particular those in Sestola, where we measured values oscillating between the very low 
$q = 6.3$ and $q$ = about 30; nor it is possible to invoke effectively in support of this hypothesis the deduction by Mc. Lennan, whose opinion was that the value of  $q$ should be deeply linked to the amount of active substances contained in the soil above which measurements are made. Strong's recent experiments lead also to think that the source of these penetrating radiations is outside the soil.

However, also if one supposes to remove, at least for the largest part, the influence of active substances contained in the soil, it is still very difficult to find the origin of the penetrating radiation which is able to ionize a gas inside a metallic vessel, and in particular the cause of the double daytime peak commonly observed. This radiation follows the same trend as the rate of decrease of the potential (Wood and Campbell, l.c.): it shows a minimum by the maximum of the ionization in free air, measured by Ebert's device, and oscillations similar to those of the unipolarity $Q={I_+}/{I_-}$. These facts roughly match, if we suppose that, at least in the open country, Ebert's device detects most of the charges contained in air, particularly when this is free from fog and there are no winds which lift big charged material particles. The daytime drop of potential and unipolarity depends
on the conditions of the air above the surface of the soil\footnote{A. Gockel, {\em Meteor. Zeit.} {\bf 25}, 9, year 1908; D. Smirnow, {\em Phis. Zeit.} 337, year 1908.}, rather than on those on the high layers of the atmosphere; thus, before formulating new hypotheses like that of Richardson\footnote{O.W. Richardson, {\em Nature} {73}, 607, year 1906.} (supported by Mc. Keon, l.c.) who, starting from the fact observed by Wood and Campbell (that $q$ oscillates like the potential drop) similarly to the concept  of Arrhenius\footnote{Arrhenius, {\em Kosmische Physik,} ii, pag. 890.}, finds the Sun being the direct cause of these oscillations of the penetrating radiation which is able to ionize a gas inside a closed container above the Earth's surface; one must consider how much the $\gamma$ radiation due to active substances contained in air affects the ionization in a closed vessel; and so we must verify first of all if also the whole amount of radioactivity induced in air, undergoes  near the soil daytime variations similar to those of the electric field in atmospheric layers close to the soil surface.

\setcounter{footnote}{0}

\include{pacini1910}

\setcounter{footnote}{0}


\include{pacini1912}


\end{document}

%% file: pacini1910.tex
%
%

\section{\BY{D. Pacini} {\em ``Penetrating radiation on the sea'',} \IN{Le Radium} {VIII}{1911} {307}, translated and commented by M. De Maria and A. De Angelis}

\vskip 1cm

\begin{center}

{\center{\bf{PENETRATING  RADIATION ON THE SEA}}}

{\center{Note by D. PACINI}}

\end{center}

\vskip 1cm
%
The penetrating radiation observed in the air over the soil surface is coming partly from the active substances in the upper layer of the Earth's surface as well as from their disintegration products, and partly from outside the soil.

The penetrating radiation originated from outside the soil is due at least in part to the transformation products of the radioactive elements of the air. 

The radiation originating from the soil changes from place to place with the changes of the soil nature, in proportion to the content of active materials; in the same place, it may be reduced with an increase on the permeability of the upper layers, for example because of rainfall, but, for a dry soil, it should remain 
approximately constant. 

The radiation originating from outside the soil varies with the atmospheric variations, being influenced by winds, rainfalls, modifications of the electric fields of the Earth; all these reasons can determine the accumulation of the active materials inside the lower layers of the air and the upper layers of the soil.

Many investigators, studying this phenomenon, have noticed considerable variations. For example, the author has found in Sestola in 1908 that the penetrating radiation entailed ionization rates from 6 to 30 ions per cm$^3$
 per second\footnote{D. Pacini, \textit{Rend. Lincei}, \textbf{18} (1909) 123.}. Similarly Mache,\footnote{H. Mache \textit{Sitzungsberichleu der K. Akad. der Wiss. Wien}, \textbf{119} (1910).} in his experiments conducted in Innsbruck from October 1st 1907 to October 15th 1908, reached the conclusion that the penetrating radiation due to the active substances in the air, or fallen from the air on the soil, may be 4 times greater than that coming the soil itself. 
 
Up to now, the results obtained show that if one measures the penetrating radiation in a place, with no influences from atmospheric variations or electric fields, where the soil is relatively rich in active substances, the importance of the soil and of the walls is predominant, with only minor variations in the series of values recorded. Conversely, if one performs the same measurement in a place, like Sestola, highly exposed to meteorological perturbations and variations of the electric potential, a larger variability of the penetrating radiation values is expected.  

However, according to Wulf and Kurz, most of the radiation observed in air near the ground would be due to active substances of the surface layer and the radiation that does not come from inside the ground would be negligible compared to the other. The conclusion of Wulf's work are:
\begin{enumerate}
\item The penetrating radiation is caused by primary radioactive substances lying in the uppermost layers of the Earth extending down to about 1 m below the surface.
\item The fraction of the radiation stemming from the atmosphere is so small that is impossible to be detected by the methods that have been used.
\item The time fluctuations in the $\gamma$ radiation can be explained by displacements of emanation-rich air masses under the surface of the Earth at larger or smaller depths.
\end{enumerate}

We see then that the results of experiments in Sestola, in Innsbruck and elsewhere are not in agreement with the conclusions of 
Wulf\footnote{Th. Wulf, \IN{Le Radium} {7} {1910} {171}.} and Kurz\footnote{K. Kurz,  \IN{Phys. Zeit.} {10} {1909}{834}.}. It is therefore necessary to do more research to see if the variable part of the penetrating radiation, that can sometimes take significant values, is really due to active substances contained in the soil, or whether  its origin should be searched, at least in part, outside the soil.

These facts suggested to conduct some experiments on media capable to absorb $\gamma$ radiation from the soil; for this purpose, I performed  the series of observations at the surface of the sea, which will now be discussed.

The experiments were made in Livorno, at the Naval Academy, 
with two Wulf's devices.\footnote{Th. Wulf, \IN{Phys. Zeit.} {10} {1909} {152}.} Such equipments, besides their low capacity (the ones I have used had capacities of 1.2 cm and 1.38 cm respectively) have the advantage to enable detecting possible losses caused by insulation faults.

\vskip 0.4cm
\begin{center}
{\center {\it {Comparison between the indications of the two devices}}}
\end{center}
\vskip 0.3cm

The two devices I used will be designated by the letters A and B.

I should note that the thickness of the walls of the device A is larger than  B. The loss $\Delta V_i$ due to isolation faults varied from a negligible amount, to 0.3 V per hour; only exceptionally it has reached 0.5 V. The first observations were made at the meteorological observatory of the Naval Academy. Denoting by $\Delta V$ the potential drop in volt per hour, we find that the mean of observations made with the device A $(\Delta V - \Delta V_i)$ is of 16.3 volt per hour, while the device B gives for $(\Delta V - \Delta V_i)$ the value of 25.4 V/h. These two numbers correspond, respectively, to 14.6 and 25.8 ions per cm$^3$ per second. From this value one must subtract the amount of ions which, as it will be discussed below, may be due to the direct action of the walls of the device; as a result I found:
\begin{eqnarray*}
n & = & 14.6-~4.7  = ~~9.9\mathrm{~ions~per~cm^3~and~per~second~with~device~ A}\\
n & = & 25.8 -11.0 = 14.8 \mathrm{ ~ions~per~cm^3~and~per~second~with~ device~B}\, .
\end{eqnarray*}

Some measurements with both devices were then performed in the garden of the Naval Academy.

The Academy is on the seaside. The building has on the land side a garden where I placed the two devices protected from the direct radiation from the Sun. These observations allowed me to assess how the two devices behave when placed in the same conditions. The experiments were continued for 10 days and measurements were made with each device every hour, from 7am to 8pm.

\begin{center}
\resizebox{0.7\columnwidth}{!}{\includegraphics{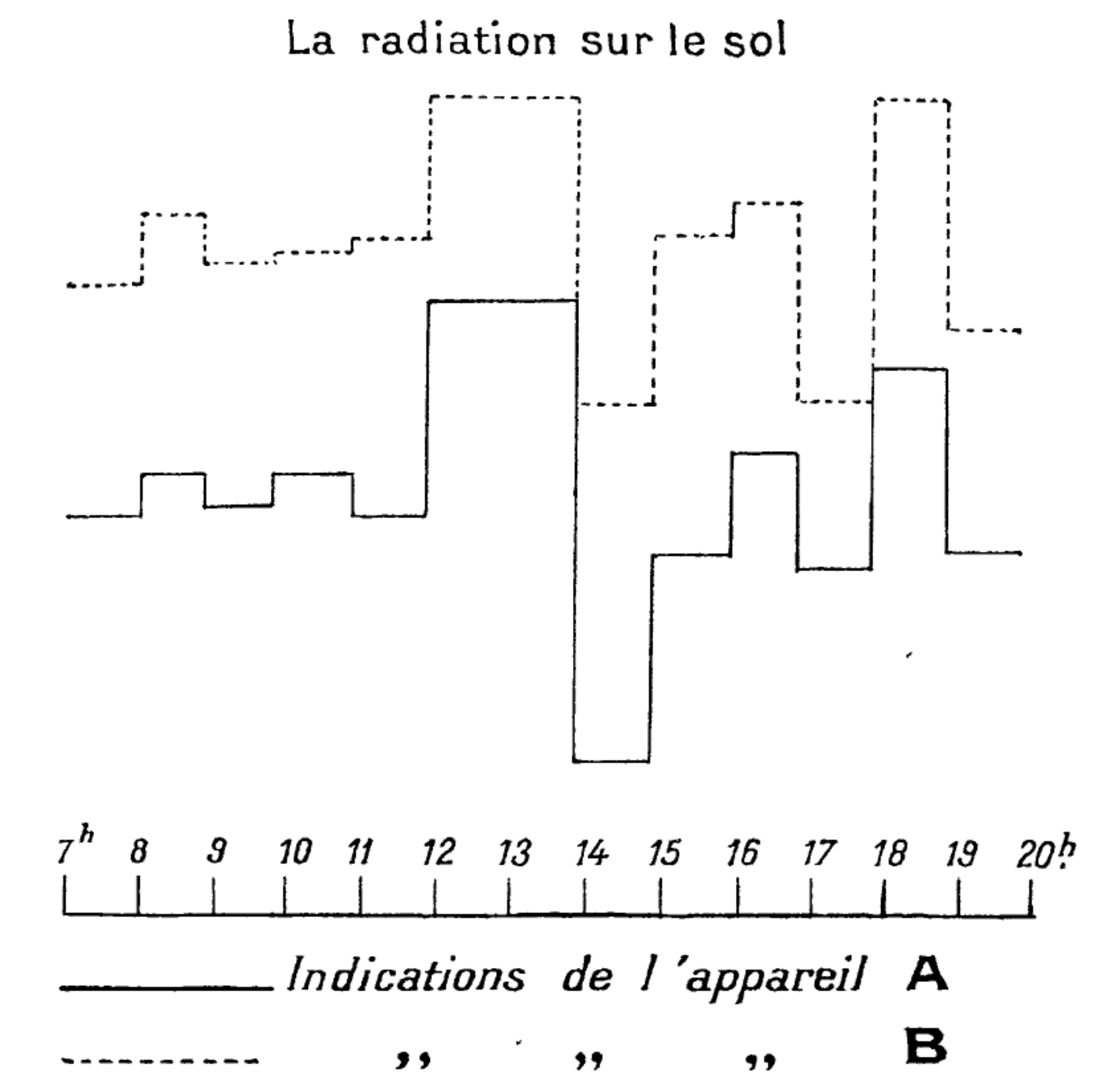} }

{Figure 1}
\end{center}

The diagram in Figure 1 displays the averages obtained taking into account all observations, and shows the time dependence of the phenomenon of penetrating radiation into this place during the 10 days in which observations were performed.
The same diagram allows us to compare the indications of the two devices operating simultaneously in the same conditions.

If we consider the complexity of the phenomenon studied, that is influenced, as we know, by the device itself, we conclude \textit{that both devices follow, with sufficient agreement, the variations of the penetrating radiation the atmosphere}.

To see how much of the observed ionization may be attributed to radiation coming from outside, it is necessary to minimize the action of external agents and to do this the two devices were placed in various locations, shielded by thick lead walls. They were exposed to the sea and on the Bracciano Lake at a distance from the shore large enough to totally exclude the action of the soil by the air layer interposed, and with a depth of the water large enough
that the action of the sea (lake) bed was negligible.
Furthermore the device A was immersed in the water of the Bracciano lake. However the absorbing layers were not successful in reducing the internal ionization below the smallest values obtained in the series of experiments on the surface of the sea.

The minimum value for $(\Delta V-\Delta V_i)$ obtained on the sea with the device A, was of $5.3$~V, which is equivalent to 4.7 ions per cm$^3$ per second. The device B measured a  minimum of $10.7$~V,  {\em{i.e.,}} 11.0 ions per cm$^3$ per second.

We can therefore say that the walls of the device A generate by themselves at most 4.7 ions per second per cm$^3$ of air contained within in device, and the walls of  device B at most 11.
The differences between the results of isolated observations and and the numbers above can be attributed to the action of penetrating $\gamma$ radiation from outside and to the action of secondary radiation generated by the first interacting with the walls of the device.

From the experiments done in the garden of the Academy, it follows that the device provides a minimum of 6.5 ions, a maximum of 16.6 and an average of 14.6 per cm$^3$ per second. Subtracting the 4.7 ions that we assume to be due to the walls, we have for the device A on the ground:

\vspace{3 mm}
\begin{tabular}{ l  l  l}
  Minimum:    & 1.8    ions per cm$^3$ and per second; & \\
  Maximum:  & 11.9 & \\
  Average:  & 6.9 &  \\
\end{tabular}
\vspace{4 mm}

\noindent and for the device B also on the ground:

\vspace{3 mm}
\begin{tabular}{ l  l l}
  Minimum:    & 3.6 ions per cm$^3$ and per second; & \\
  Maximum:  & 25.8  & \\
  Average:  & 14.1 &  \\
\end{tabular}
\vspace{3 mm}

We see that for the first unit the oscillations of the values of the penetrating radiation over the ground are such that the minimum is $26\%$ of the average and the maximum is $172\%$
of the same average.

For the second device, the minimum is $25\%$ of the average value and the maximum is $169\% $.

We can thus conclude that \textit{both devices operating simultaneously on the ground showed some oscillations in the values of the penetrating radiation that are in a great approximation of the same magnitude}.

\vskip 0.4cm
\begin{center}
{\center {\it {Observations on the sea}}}
\end{center}
\vskip 0.3cm

With one device (device B), we continued our observations in the garden of the Academy. The other (A) was put on a dinghy of the Navy and protected from direct radiation of the Sun. The dinghy was anchored at more than 300 meters from shore, a distance capable to reduce to a fraction 
of less than 4 per 100\footnote{K. Kurz, \IN{Phys. Zeit.} {10} {1909} {832}.} the radiation coming from the mainland; 
as at the point chosen for anchoring the depth of the sea was beyond 4 meters, the radiation from the seabed was completely 
absorbed\footnote{Wright, {\em {Phil. Mag.,}} February 1909.}.

This first series of observations spanned nine days between August 3 and August 19. Because of unavoidable circumstances, bad weather
in particular, we had to suspend the tests for a few days.

The result of this series of observations was that the device A measured a minimum of 4.7 ions, a maximum of 15.2 and an average of 8.9.

Since the minimum of 4.7 is compatible with the radiation from the walls, we find that:

%
%
%
%

\[
a)\mathrm{~on~the~sea~(device~A)}\left\{\begin{array}{l c c r}
	\mathrm{minimum ~very~low} & ~~~~& ~~~~~\\
	\mathrm{maximum ~of~10.5~ions} & ~~~~& ~~~~~\\
	\mathrm{average~of~4.2}& ~~~~& ~~~~~
	   \end{array}\right.
\]
\noindent while
\[
\mathrm{~on~the~ground~(device~A)}\left\{\begin{array}{l c c r}
	\mathrm{minimum ~1.8~ ions} & ~~~~& ~~~~~\\
	\mathrm{maximum ~of~11.9~ions} & ~~~~& ~~~~~\\
	\mathrm{average~of~6.6~ions}& ~~~~& ~~~~~
	   \end{array}\right.
\]

\noindent where we took into account all the observations made on land with the device A. 
In average the unit A measures  on the sea 2.4 ions less than on the land.

We can compare the the oscillations observed on the sea and those found on ground during the same period with the device B. 
Device B measured in the period from 3 August 19: 

\[
a')\mathrm{~device~B~at~ground~}\left\{\begin{array}{l c c r}
	\mathrm{minimum ~7.8~ ions} & ~~~~& ~~~~~\\
	\mathrm{maximum ~of~19.9~ions} & ~~~~& ~~~~~\\
	\mathrm{average~of~12.8~ions}& ~~~~& ~~~~~
	   \end{array}\right.
\]

Considering together the data $a$ and $a'$ we see that the maximum value obtained on the sea is $ 255\%$ of the average, and the maximum value obtained simultaneously on land is $155\%$
of the average. 

Conclusion: while both devices showed on the ground the same oscillations of the penetrating radiation, now the device A shows oscillations that are clearly larger than those measured in the same time on the ground using the device B. 

From August 20 to August 26 the position of the devices was exchanged: B was transferred on the sea and A in the garden of the Academy. 
The device B on the sea measured a minimum of 11.0 ions, a maximum of 20 and an average of 19.4, and assuming that 11 ions are due to the device itself, we find that: 

\[
b)\mathrm{~on~the~sea~with~the~device~B~}\left\{\begin{array}{l c c r}
	\mathrm{minimum ~very~low} & ~~~~& ~~~~~\\
	\mathrm{maximum ~of~15~ions} & ~~~~& ~~~~~\\
	\mathrm{average~of~8.4}& ~~~~& ~~~~~
	   \end{array}\right.
\]
\noindent while
\[
a)\mathrm{~on~the~ground~with~the~device~B}\left\{\begin{array}{l c c r}
	\mathrm{minimum ~of~3.6} & ~~~~& ~~~~~\\
	\mathrm{maximum ~of~23.8~ions} & ~~~~& ~~~~~\\
	\mathrm{average~of~13.4~ions}& ~~~~& ~~~~~
	   \end{array}\right.
\]

\noindent taking into account again of all the observations made in the land with B. 

In average the unit B measures on the land 5 ions more than on the sea.

For data on land from August 20 to August 26: 

\[
b')\mathrm{~on~the~ground}\left\{\begin{array}{l c c r}
	\mathrm{minimum 3.0} & ~~~~& ~~~~~\\
	\mathrm{maximum ~of~10.0~ions} & ~~~~& ~~~~~\\
	\mathrm{average~of~6.4}& ~~~~& ~~~~~
	   \end{array}\right.	   
\]

Considering all the data $b$ and $b'$ we see that the maximum obtained at sea is $178\%$
of the average value, the maximum obtained in the same time on the land being $172\%$
of the corresponding average. This second series of experiments indicates that the penetrating radiation on the sea undergoes oscillations that are at least of the same order of magnitude as on the land.

About \textit{the evolution of the phenomenon on the sea surface and on the land,} I notice that, when the two devices operate on the land under the same conditions, Figure 1 reveals for both the same trend of the penetrating radiation during the ten days of observation, and it would be of great interest to compare the trend on land and sea; it would be sufficient to compare the series of observations made on the sea with the series on the land during the same period. But it is clear that in order to show the existence of a possible correlation, it is necessary to do a long series of measurements in order to reduce the errors due to unavoidable local influences, determined by many occasional causes, to a minimum. The winds that can transport variable amounts of active material in the vicinity of the devices, the location of the instruments, the holders and the shelters can be sources of error. On the sea, at a suitable distance from the shore, these sources of error can be reduced; in any case,  to make a significant comparison, a period of time longer than that I dedicated to the experiment would be needed.

However, I plotted the results of the observations made on the sea with the device A, and on the land with the device B, for the first and longest series of measurements made simultaneously at the sea surface and on land of from 3 to 19 August. The diagram in Figure 2 was constructed taking in account the mean values at every hour. Apart from the very high value obtained between 7:00 and 8:00 am on the land with device B, a correlation between the two trends can be perceived. But I repeat that for deciding on this question, a longer series of observations would be necessary.

\begin{center}
\resizebox{0.7\columnwidth}{!}{\includegraphics{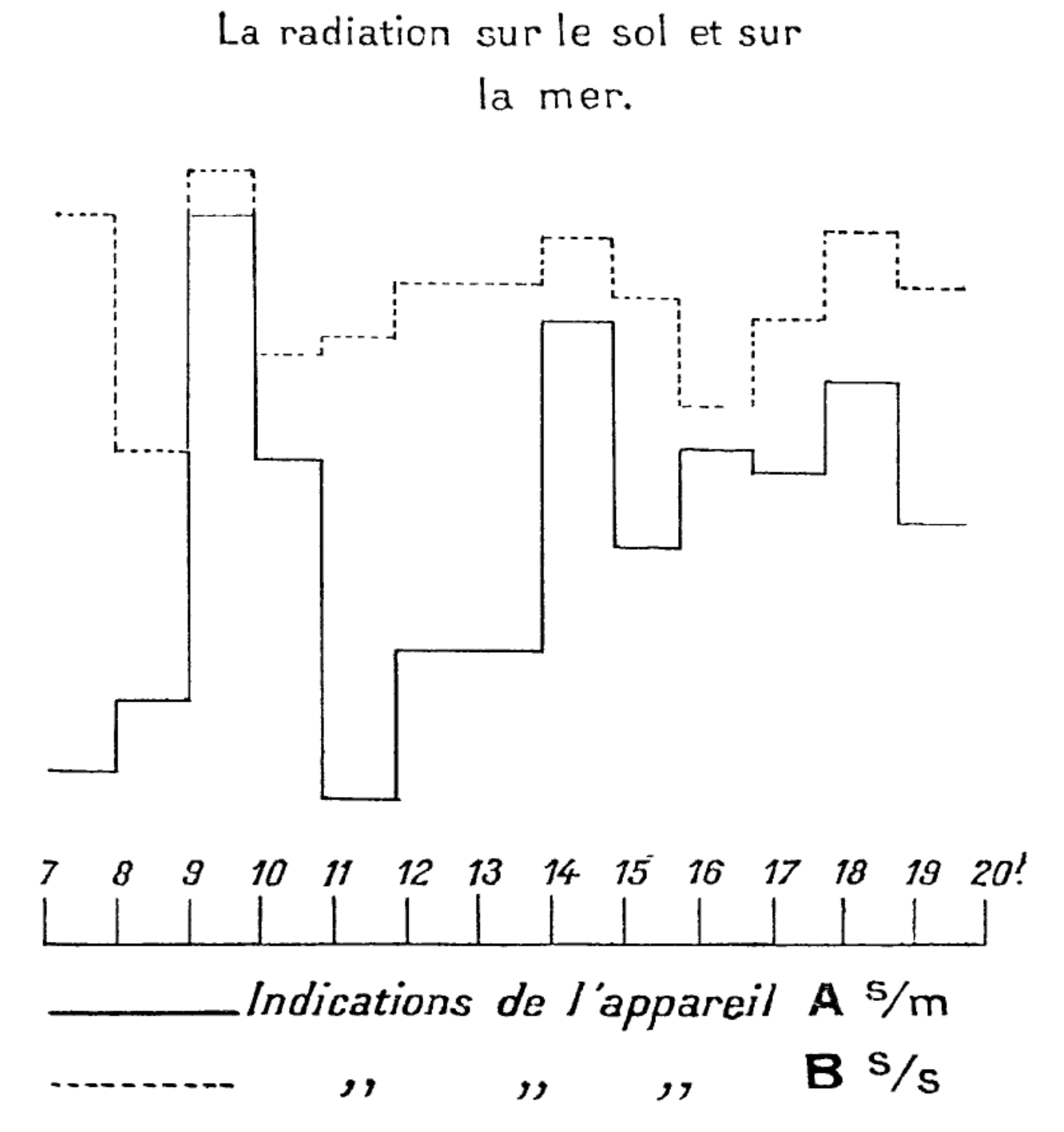} }

{Figure 2}
\end{center}

We now want to suggest a possible explanation of the noticeable oscillations observed at the sea surface. The radiation observed on the sea, 
as far as we know, can not be attributed to the active materials in the air and to the products of disintegration of the radioactive emanation contained in the sea, products whose action on the surface could, as noted by Kurz, be spread in a more effective way due to the effect of the sea waves. We found however from our observations, that there is no relationship between the state of the sea and the value of the penetrating radiation, so \textit{we do not have any evidence to attribute oscillations of this magnitude to the decay products of radioactive emissions released into the sea by the effect of waves.} 

About this result, I should recall another result that I have obtained in the Gulf of Genova\footnote{D. Pacini,  \IN{Nuovo Cim.}{15}{1908}{18}.}, whre, studying the ionization on the open sea, I found that often such ionization was relatively low in rough sea conditions.

During the days in which the experiment was performed, the prevailing winds blew from the west, that is to say, from the sea; there were anyway also observations made under winds from the coast.

An examination of the results shows that with low or very low winds from the coast, the values of the penetrating radiation tend to be below the average, and that with moderate winds from the sea there is also a predominance of values below the average.

Note that no observations were made in presence of strong winds. Moreover, when observations were suspended because of atmospheric disturbances,  violent sea winds were often blowing.
Sometimes the observations began few hours or even few moments after the wind had stopped, and one can assume in such conditions that the physical properties of air, at least at the sea surface, must be influenced for a certain amount of time by the conditions determined by the winds that previously had blown so violently. A long series of measures would be required to determine if on the sea, at a certain distance from shore, a difference of action related to the difference of the wind direction and speed exists.


\vskip 0.3cm
\begin{center}
{\center {\it {Summary}}}
\end{center}
\vskip 0.3cm

\begin{enumerate}
\item The number of ions due to penetrating radiation on the sea is estimated to be 2/3 of that on the ground, and  consistent values for this ratio have been measured using two different devices.
\item Under the experimental conditions described in this work and according to the measurements provided by the device B, whose walls have a thickness suitable to allow the passage of the largest part of highly penetrating radiation,  the number of ions generated  per cm$^3 $
per second by this radiation at the sea surface, is smaller by 5 ions with respect to the average value observed on land. We assumed for the charge of the ions the new value $e = 4.65 \times 10^{-10}$ U.E.S.
\item The penetrating radiation on the sea at a distance of more than 500 meters from shore with water depths larger than 4 meters, under conditions that allow to neglect the radiation from the soil, undergoes oscillations that are at least of the same order of magnitude than observed at the same time on the ground.
\item As no relationship appears to exist between the state of the sea and the action of gamma rays, we cannot attribute the oscillations of the radiation level to decay products of emanations that could possibly be released from the water with larger intensity, according to the intensity of the sea waves related to storms.
\item During the experiments, the winds blowing were generally weak and came from the sea; the data collected do not allow establishing whether a
relationship exists between the penetrating radiation  and the wind's speed and direction.
\end{enumerate}
These results agree with those  obtained by the author in Sestola and later by  Mache in Innsbruck: in the air at the sea surface, as well as  on the rocks of the Apennines near Modena,
I found that the penetrating radiation may undergo large oscillations from relatively high values to so small values that they could be attributed almost entirely to the action of the walls of the device. Large oscillations occur near the rocks of the mountains, on the air above the sea, and on the ground, where the ionization produced by radiation is larger because of the direct action of active substances from the soil.

The number of ions due to penetrating radiation on the sea being the 2/3 of those on land, we can explain the relatively high value of the ionization of the free atmosphere at the surface 
of the sea\footnote{D. Pacini, \textit{Nuovo Cim.,} \textbf{15} (1908) 18.}. An important question remains open, since our knowledge of the quantity of active substances in sea water and air does not allow us to explain the large values found for the penetrating radiation on the sea,\footnote{A.S. Eve, \textit{Terr. Magn. and Atm. Elect.,} 1910 -- D. Pacini, \textit{Nuovo Cimento}, (1910) 449.} nor on the mainland (Gockel's balloon 
measurements\footnote{A. Goeckel, \textit{Phis. Zeit.,} (1910) 280.} and Wulf's measurements on the Eiffel Tower\footnote{A. Wulf, \textit{Phis. Zeit.,} (1910) 814.}) at a sufficient height that one can neglect the action of the active substances form the soil. Anyway such results seem to indicate that {\em
a substantial part of the penetrating radiation in the air, especially the one that is subject to significant oscillations, has an origin independent of the direct action of active substances in the upper layers the Earth's crust.}


\vskip 5mm

 $\mathrm{\left[Manuscript\: received\: April\: 2,\: 1911\right].}$
 
 \hskip 5cm $\mathrm{\left[Translated\: by \:L.\: Bloch\right].}$


%% file: pacini1912.tex
\section{\BY{D. Pacini} {\em ``Penetrating radiation at the surface of and in water'',} \IN{Nuovo Cim.} {VI/3} {1912} {93}, translated and commented by A. De Angelis}

\vskip 1cm

\begin{center}
{\center{\bf{PENETRATING  RADIATION AT THE SURFACE OF\\AND IN WATER}}}
\end{center}

\vskip 1cm

Observations that were made on the sea during the year 1910\footnote{D. Pacini. Ann. dell'Uff. Centr. Meteor. Vol. XXXII, parte I, 1910.\\- Le Radium, T. VIII, pag. 307, 1911.} led me
to conclude that a significant proportion of the pervasive radiation
that is found in air had an origin
that was independent of direct action of the active substances
in the upper layers of the Earth's surface.

Here, I will report on further experiments that support
this conclusion.

The results that were previously  obtained indicated that a source of ionization existed on
the sea surface, where  possible effects from the soil are small, that had such an intensity that 
could not
be explained on the basis of the known distribution
of radioactive substances in water and in air.

Indeed, as shown by Eve\footnote{A.S. Eve. Phil. Mag., 1911.}, one can easily calculate
 the expected ionizing action at the surface of the 
sea, 
due to
$\gamma$ radiation that is emitted by active particles in air.

Let:
\begin{itemize}
\item[-] $Q$  be the equivalent radiation in Ra. C per cm$^3$ in the atmosphere, expressed
in grams of Radium in radioactive equilibrium, $Q = 8 \times 10^{-17}$.
\item[-]  $K$ be the number of ions that is generated per cm$^3$ per second from one gram
of Radium at a distance of 1~cm:  $K = 3.4 \times 10^9$ for air enclosed
in an Aluminum electroscope;  $K = 3.1 \times 10^9$ for free air.
\item[-]  $\lambda$ be the absorption coefficient of $\gamma$ rays in air =
0.000044.
\item[-]  $r$ be the distance from the point at which we consider the action.
\end{itemize} Then, the number $q$ of ions due to the $\gamma$ rays  of  Radio C in
air will be expressed as:
\begin{eqnarray*}
q & = & 2 \pi K Q \int_0^\infty \frac{r^2e^{-\lambda r}}{r^2} dr\\
q & = & 2 \pi \frac{K Q}{\lambda} = 0.035 \, .
\end{eqnarray*}

We should now  take into account the effect of the active products of Thorium, but there is no precise input
to complete the calculation. Eve assumes that, due to the effect
of the $\gamma$ radiation that is emitted by the products of 
Thorium, 0.025 ions per cm$^3$ per second are generated; which gives a   
total of 0.06 for the  ions in air.
In this calculation it is assumed that the air above the
sea surface has the same radioactive composition as the
air above the ground, 
while indeed at some distance from the coast, the content of radioactive emanation of the air above the sea is smaller with respect 
to the air above the soil, especially
regarding Thorium.

For the contribution from sea water, the calculation is
also easy, knowing from the Joly\footnote{Joly. Phyl. Mag., September 1909.} experiments that
the equivalent in Radio is $Q '= 1.1 \times 10^{-14}$; the absorption coefficient $\lambda'$
is  immediately obtained, using the relationship between
that coefficient  and the density:
\[ \frac{\lambda'}{\rho} = 0.034 \, \footnote{Mc. Clelland. Phil. Mag., July 1904.} \, . \]
Thus, we obtain the value $q = 0.006$ for  the sea.

To the value $q = 0.066$ we must add the effect of
secondary radiation emitted from the walls of the container;
we can assume that this will increase the effect that one has
in open air by 20\%, so we arrive at an estimate of a total ionization
that is on the order of 0.1 ions per cubic centimeter.

The measurements that I made on the sea had nonetheless
provided for values for $q$ that were, on average, significantly larger 
than those predicted by theory.
As an example, I take the
indications from the apparatus A \footnote{See Pacini, op. cit.}, which had walls 
that were 1.5 mm thick,
to exclude the vast majority of $\beta$ radiation.
Onboard
a boat with a surface of about 4 m$^2,$ this device provided an average measurement of
8.9 ions on the sea, with  a minimum of 4.7; in the hypothesis, supported by the results obtained so far, 
in which the minimum of 4.7 ions
can be ascribed entirely to the residual ionization, one is left with
an average of 4.2 ions, from which subtracting the action of
secondary radiation, we obtain the value:
\[ q = 3.4 \, \rm{ions} \]
 due to the penetrating radiation over the sea at a distance greater than 300
meters from the coast.

Subsequently, in May 1911,  
Simpson and Wright\footnote{G. C. Simpson and C. S. Wright:  \emph{Atmospheric Electricity over
the Ocean.} Proceed. of the Rovedo Soc. Vol. 85, p. 175, 1911.} published a note, reporting observations of
atmospheric electricity aboard the ``Terra Nova'' in a journey
from England to New Zealand, following the Antarctic expedition of
Captain Scott. 
The authors observed onboard
their ship, on average, a value of approximately 6 ions 
for the penetrating radiation; however, they found during several hours
values of approximately 9 ions after the ship
had left the coast -- an increase of 3 ions with respect to the
average value of $q.$ The minimum  value that was obtained for $q$ was
4 ions.

The results by Simpson and Wright confirm
that even outside of the direct action of soil it is possible to observe
considerable fluctuations in the values of penetrating radiation. The results 
of the experiments on which I will now report also seem to indicate the presence of well-measurable 
effects of 
penetrating radiation in air, over an
absorbing medium.

We shall see that by immersing the measuring device in
water, one can further lower the penetrating radiation that is observed at the surface
of a sea or lake below its average
value.

The A device, already used in the experiments above,
was enclosed in a copper box to be able to immerse it in water. The experiments were performed again
at the Naval Academy of Livorno, precisely
in the same place where the measurements of the previous year had been made. 

The apparatus was put onboard the same boat,
which was pegged at more than 300 meters from the coast, over 8-m-deep water.
Between June 24 and June 30, measurements were made
with the apparatus on the surface   and immersed at a depth of 3 meters.

Here are the results of these observations, each of which
had a duration of approximately 3 hours:

When  the instrument was at the surface, the loss per hour, measured in volts, was:
\[ 13.2 - 12.2 -  12.1 - 12.6 - 12.5 - 13.5 - 12.1 - 12.7\]
(average of 12.6, corresponding to 11 ions per cm$^3$ per second).

With the instrument immersed:
\[ 10.2 - 10.3 - 10.3 - 10.1 - 10.0 - 10.6 - 10.6 \]
(average of 10.3, corresponding to 8.9 ions per cm$^3$ per second).

The difference between these two values is 2.1 ions.

The boat was the same as the one that was used for the measurements in
which the minimum value of 4.7 ions was established, and because
 it had always been kept under the same conditions --  {\em{i.e.,}}, either on the
sea or suspended over the sea from the quay -- we are convinced that the boat did not contain
active materials other than  those that came from the air or 
sea. In the hours in which measurements were not made, the measuring apparatus
was kept charged, always in the same room, and the dispersion
of electricity was strictly constant.

With the same apparatus, measurements were also made at the
Lake of Bracciano. At 350 meters from the shore,  I measured 
$q = 12.4 q$ at the surface, while at a depth of 3 meters
(at a place where the bed 
was over 7 m deeper), the result was
$q = 10.2.$ Thus, the difference in the two values of $q$ was 
2.2 ions.

With an absorption coefficient of 0.034  for water, it is 
easy to deduce from the known equation $I/I_0 = e^{-\lambda d}$, where $d$ is the
thickness of the matter crossed, that, in the conditions of my
experiments, the activities of the bed and of the surface were both
negligible.

The water temperature was, on average, a few tenths of a degree
lower than the air above, and working under
airtight conditions, the number of ions created in the internal space
 varies only due to changes in radiation. From the measured  differences of
2.1 and 2.2, by subtracting 20\% due to secondary radiation,
these numbers become:
\begin{center}
\begin{tabular}{l}
1.7 ions for the sea\\
1.8 ions for the Lake of Bracciano.
\end{tabular}
\end{center}

Is this decrease in the value of $q,$ moving from observation at
the surface to a survey of the interior of the water,
 due to external actions or rather to a variation in the residual
ionization of the container in the transition from air to water?

We know nothing for certain about the origin of the
residual ionization for air  that is trapped in a metallic container.

The causes that might generate the
residual ionization  
are intrinsic activity, or radioactive impurities of the
metal, and possible spontaneous ionization  of the enclosed gas\footnote{G.C. Simpson and C.S. Wright (op. cit.).}.

 Under the conditions in which 
these experiments are conducted, it is unlikely that metals, with the exception of
lead, contain radioactive impurities. Furthermore, in a
long series of observations that were made earlier with the same apparatus, an increase in dispersion
that could be ascribed to impurities was never observed.

In the case of activity that is due to the metal or 
spontaneous emission of electrons due to disintegration of the gas
that is encapsulated in the device, one cannot see any reason for a variation
of these causes of ionization, given the changing
conditions of  the isolated instrument between the surface
and the depth.

The explanation appears to be that
 due to the absorbing power  of water and the minimum amount
of radioactive substances in the sea, absorption of $\gamma$ radiation
coming from the outside happens indeed, when the apparatus is immersed.

It is natural, as already pointed out\footnote{D. Pacini (op. cit.).}, 
to look for the origin of such ionization of  air due to penetrating radiation, not directly dependent on 
active substances in the soil, 
in an accumulation of radioactive material that released into the atmosphere around the
site  of observation.

Simpson and Wright also attribute
an increase of three ions to this cause, in their measurement of normal ionization on the sea.
According to these authors, the active particles could have
deposited onto the ship from the air when the ship was near the coast.

If we assume that the products are spread evenly in the atmosphere at altitudes of up to 5 km, and that they
are rapidly deposited on the surface
of the Earth from the air,  from Eve's  data one deduces a 
layer of Ra. C that is equivalent to  $4 \times 10^{- 11}$ of
Radium per cm$^3$ in equilibrium.  This would generate 1.8 ions per cm$^3$ per second in the air at a height of 1 meter.

In the case of my experiments, one can neglect the action of
active particles that deposit onto the water, because they should go into solution quickly, due to  wave motion.
We can get a sense of what the effect would be of an
active substance that is deposited onto the boat used here, that has a surface of approximately 4
m$^2.$ Suppose that the Ra. C that is deposited onto the
boat acts on the device (which is located in the center, above
a table, at the height of the edge) as if it were distributed evenly
on the surface of a half-sphere whose radius is 80 cm, in an amount $Q$ that is equivalent to 
$4 \times 10^{-11}$ grams of Ra per cm$^2$. The number of ions that is generated by the full radioactive
deposit in 1 cm$^3$  of air in the
center of the hemisphere would be expressed as
\[ q  =  \frac{K Q}{r^2} e^{-\lambda r}2 \pi r^2= 0.8 \, \rm{ions} \, . \]
Assuming that the products of Thorium are responsible for 0.5 ions in this
case, we would have a total of 1.3 ions. This calculation gives us
a value smaller than that observed, but the effect is nevertheless
well measurable.

A rapid reduction in the active products of the atmosphere
could occur for large values of the Earth's field, especially
in the case of rain. The observations that have been made so far
on the behavior of penetrating radiation during
rain are not quite in agreement, and they are not enough
to establish the existence of an
action in the sense mentioned above.

Free-balloon experiments have been performed recently\footnote{A. Goekel. Phys. Zeit., p. 595, 1911 and V.F. Hess. Phys. Zeit., p. 998, 1911.} 
on penetrating radiation in the upper atmosphere. 
Although they cannot be considered
conclusive with regard  to the study of the radiation that penetrates
at a certain height above the ground, these observations, however, could have
shown that where, according to the law of absorption from air
(recently verified by Hess), the action of
active substances of the soil is negligible, there is still a large quantity of penetrating radiation. This result has
spurred Gockel and Hess to repeat what the author of the present paper concluded from the first observations that were made on the sea and
what appears from the results of the work described in this Note:
that \emph{a sizable cause of ionization exists in the atmosphere, originating from penetrating radiation, 
independent of the 
direct action of radioactive substances in the soil.}

%
%
%
%

%% file: crrnc_new.bbl
\begin{thebibliography}{999}
\bibitem{wiki} Unless otherwise specified, the information in the short biographical footnotes is taken from \cite{treccani}; minor complements come from Wikipedia, the free encyclopedia
\end{thebibliography}
